\documentclass{jmfs}

\graphicspath{ {./fig/} }

\newcommand{\maintitle}
{Temporally-smooth Antialiasing and Lens Distortion with Rasterization Map}
\newcommand{\license}
{https://creativecommons.org/licenses/by-nc-nd/3.0/}

\hypersetup{
	pdftitle={\maintitle},
	pdfauthor={Jakub Maksymilian Fober},
	pdfsubject={Curvilinear Rasterization and Antialiasing},
	pdfkeywords={
		antialiasing;
		rasterization;
		lens matched shading;
		curvilinear perspective;
		spherical perspective;
		fisheye;
		graphics hardware
	},
	hidelinks
}

\begin{document}


\title{\maintitle}
\author{Jakub Maximilian Fober
	\thanks{
		\href{mailto:talk@maxfober.space?subject=\maintitle&cc=jakub.m.fober@pm.me}
		{talk@maxfober.space}
	\newline
		\href{https://maxfober.space}
		{\hspace{1.37em}$^*$https://maxfober.space}
	\newline
		\href{https://orcid.org/0000-0003-0414-4223}
		{\hspace{1.37em}$^*$https://orcid.org/0000-0003-0414-4223}
	}
}
\date{\textit{\today}}

\label{pg:title}
\maketitle

\begin{figure}[h]
	\centering
	\begin{subfigure}{100pt}
		\includegraphics{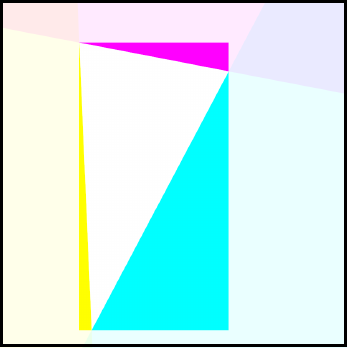}
		\caption{Half-space}
	\end{subfigure}\hspace{16.2pt}
	\begin{subfigure}{100pt}
		\includegraphics{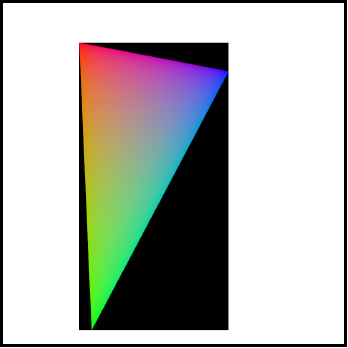}
		\caption{Barycentric}
	\end{subfigure}\hspace{16.2pt}
	\begin{subfigure}{100pt}
		\includegraphics{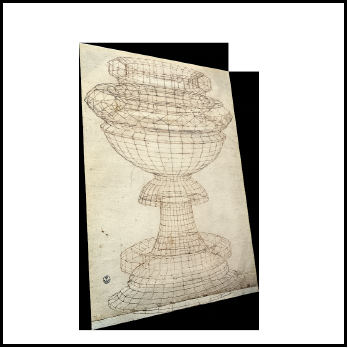}
		\caption{Textured}
	\end{subfigure}
	\caption[Rasterization example]{Rasterization example in rectilinear perspective with RMAA.}
	\label{fig:front image}
\end{figure}

\begin{abstract}
	Current GPU rasterization procedure is limited to narrow views in rectilinear perspective.
	While industries demand curvilinear perspective in wide-angle views, like \emph{Virtual Reality} and \emph{Virtual Film Production} industry.
	This paper delivers new rasterization method using industry-standard \emph{STMaps}. Additionally new antialiasing rasterization method is proposed, which outperforms MSAA in both quality and performance. It is an improvement upon previous solutions found in paper \emph{Perspective picture from Visual Sphere} by yours truly.
\end{abstract}

\vfill
\hyperlink{license}{\includegraphics{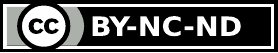}}
\pagebreak


\section{Introduction} 
\label{sec:intro}

This work provides an improvement upon \emph{Perspective picture from Visual Sphere. A new approach to image rasterization} paper.\supercite{Fober2020VisualSphere}
It simplifies rasterization algorithm and presents new improved anti-aliasing technique applicable to rectilinear 3D pipeline.
Presented solution for non-linear and aliasing-free rasterization comes in front of great demand in industries like \emph{Virtual Film Production}, \emph{Virtual Reality}, \emph{In-camera Special-Effects}, \emph{Lens-matched Real-time Graphics} and more.
It is highly parallel solution applicable for graphics hardware integration.

Previously to achieve curvilinear rasterization and aliasing-free picture for real-time graphics, extensive post processing had to be performed.
Such result was achieved through multi-view compositing, post-processing and/or multi-sampling.
Here final pixel is produced by the rasterization algorithm with no post-processing.
The product is a real-time picture with unlimited distortion, free of aliasing artifacts.

Current Computer Graphics (CG) pipeline has fixed rasterization function, which interprets polygon-data into pixels of the screen.
Upgrade to non-linear and aliasing-free imagery requires change in the hardware.

\paragraph{What you will find} in this paper is a rasterization-stage anti-aliasing technique, merging procedure for a non-binary coverage mask with fragment data. Rasterization algorithm for curvilinear projections. Algorithm for generating \emph{STMaps} in Universal Perspective and lens-distortions. Conversion algorithms and naming convention for rasterization maps.

\subsection[Paper goals]{Goals of the paper}
\label{sub:goals}

This paper aims at introducing curvilinear rasterization in an industry-friendly manner, utilizing currently used exchange formats. A contrary approach to the previous work,\supercite{Fober2020VisualSphere} which defined its own standards and formats.
Presented algorithms are optimized, ready to entry implementation procedure for next-generation hardware and software.

\subsection[Naming convention]{Document naming convention}
\label{sub:conventions}

This paper uses left-handed coordinate system, therefore cross-product is left-hand oriented.
Counter-clockwise polygons are considered front-facing. Vectors are presented natively column-oriented with matrix's vectors arranged in rows. Matrix multiplications are in column-major order, resulting in a column-vector. Vector enclosed by double-bars \textquote{$\Vert$} represent normalization function or a unit-vector, while single bar enclosure \textquote{|} represent vector's magnitude or length.
Centered dot \textquote{$\cdot$} denotes vector dot-product, and \textquote{$\cross$} a cross-product (when applicable). Vectors without dot or cross sign are multiplied component-wise and form another vector. Values enclosed by square brackets separated by blank-space denote vectors, while separated by comma, an interval.
QED symbol \textquote{\qedsymbol} marks final result and output.

\vfill
\begin{figure}[h]
	\centering
	\begin{subfigure}[t]{172.8pt}
		\centering
		\includegraphics[width=153.6pt]{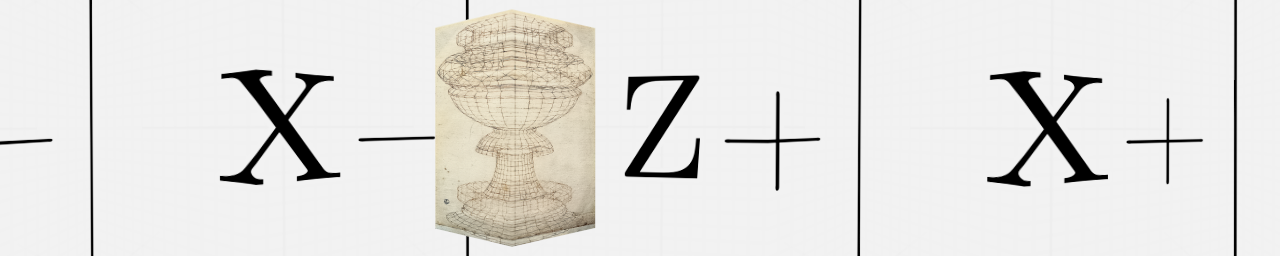}
		\caption{Five-screen array render, of $\Omega^v=60\degree$}
	\end{subfigure}\hfill
	\begin{subfigure}[t]{172.8pt}
		\centering
		\includegraphics[width=153.6pt]{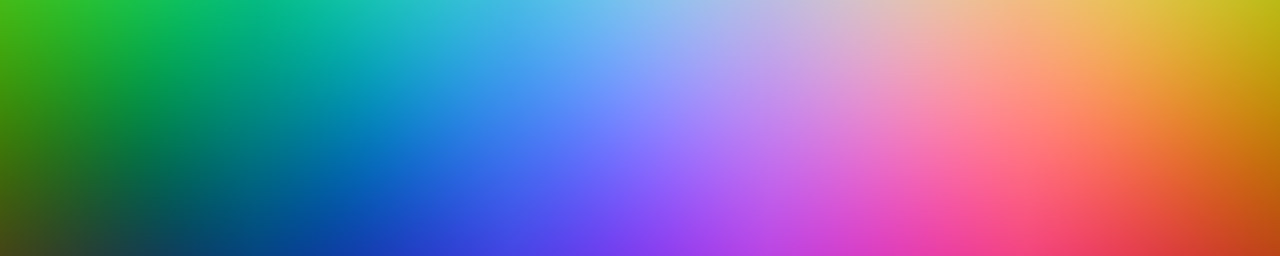}
		\caption{Five-screen array \emph{Perspective Map}}
	\end{subfigure}
	\\[1em]
	\begin{subfigure}{172.8pt}
		\centering
		\includegraphics[width=172.8pt]{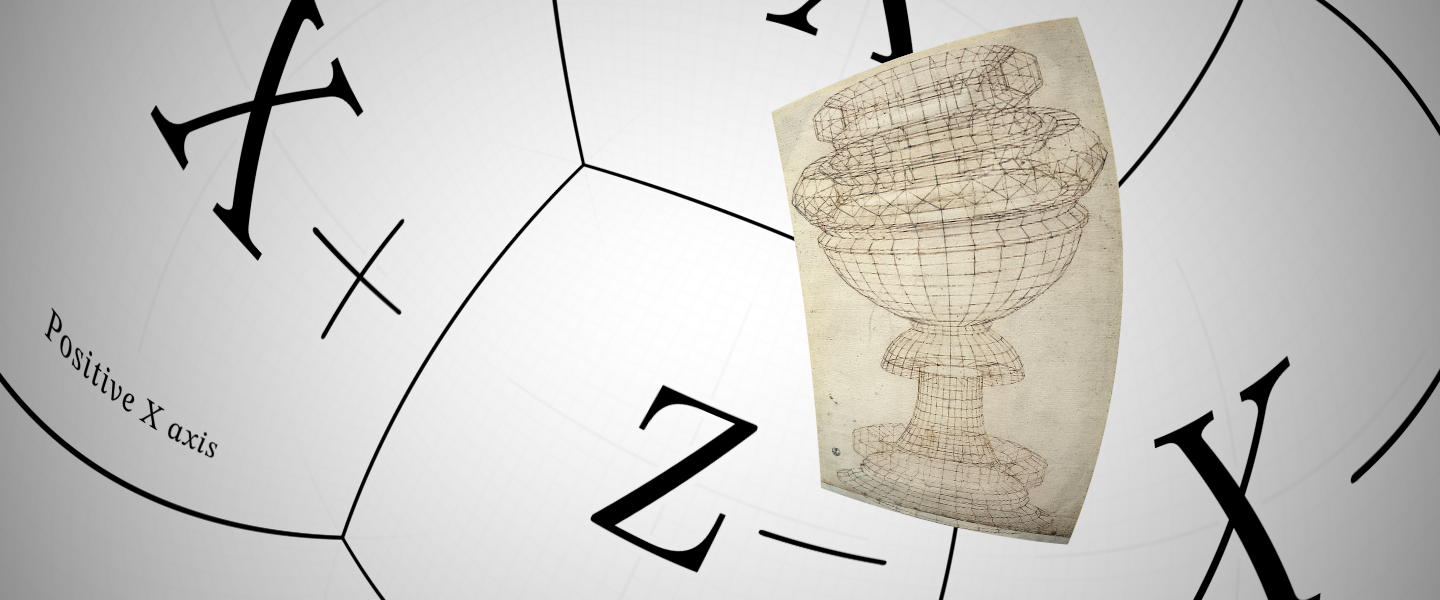}
		\caption{Fisheye render with vignette, of $\Omega=270\degree$, $k=0$, $l=0.75$, $s=98\%$}
	\end{subfigure}\hfill
	\begin{subfigure}{172.8pt}
		\centering
		\includegraphics[width=172.8pt]{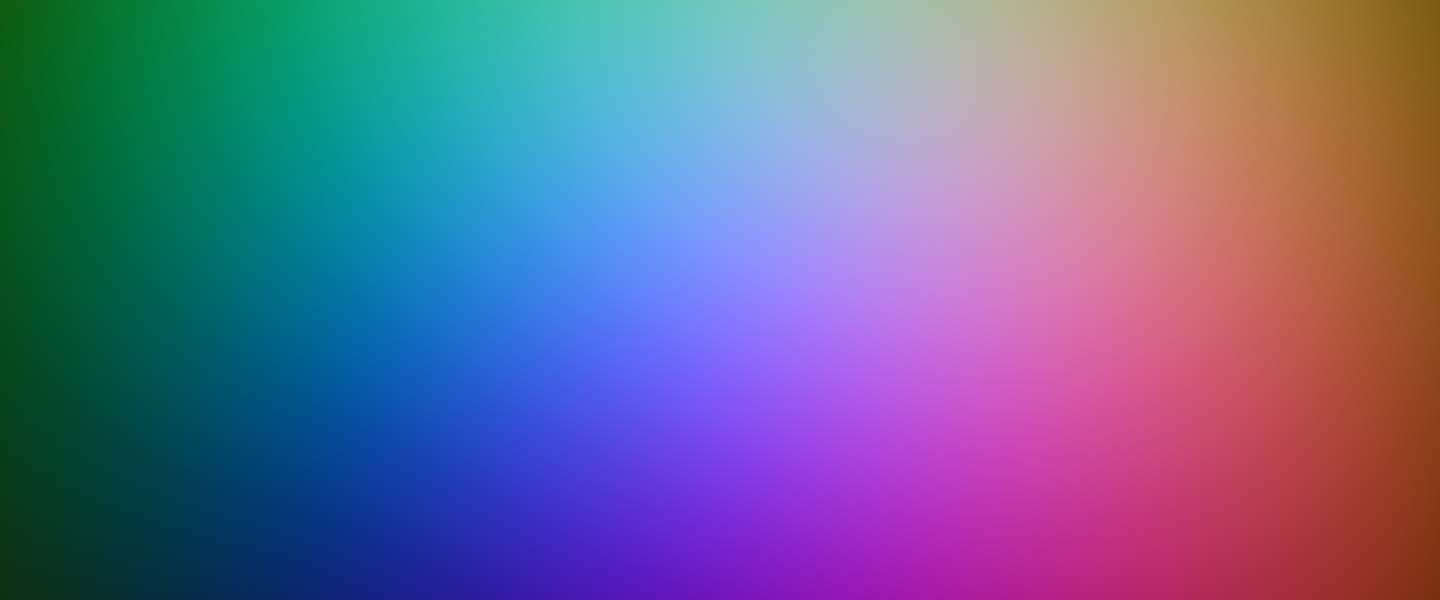}
		\caption{\emph{Perspective Map} with vignette, of $\Omega=270\degree$, $k=0$, $l=0.75$, $s=98\%$}
	\end{subfigure}
	\\[1em]
	\begin{subfigure}[t]{172.8pt}
		\centering
		\includegraphics[width=153.6pt]{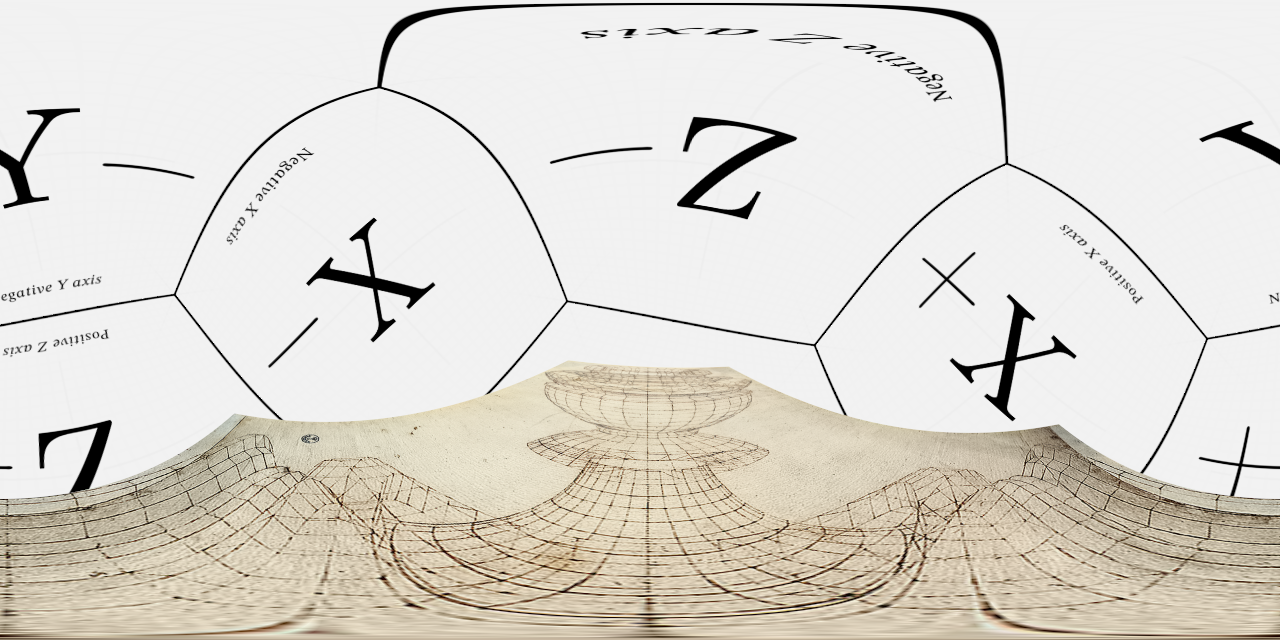}
		\caption{Equirectangular render, of $\Omega=360\degree$}
	\end{subfigure}\hfill
	\begin{subfigure}[t]{172.8pt}
		\centering
		\includegraphics[width=153.6pt]{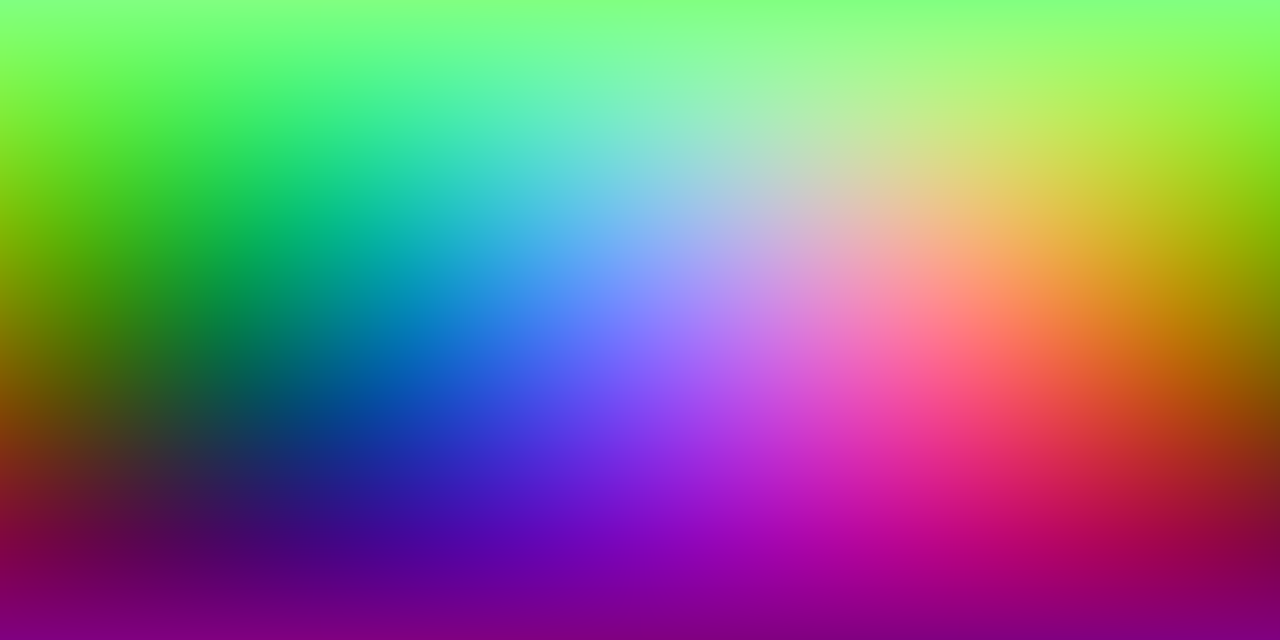}
		\caption{Equirectangular \emph{Perspective Map}}
	\end{subfigure}

	\caption[Perspective Map rasterization examples]{Rasterization examples of polygon-billboard in \emph{Perspective Map} RMAA.}
	\label{fig:rasterization perspective map}
\end{figure}
\vfill


\pagebreak

\phantomsection
\pdfbookmark[1]{Table of contents}{sec:contents} 
\tableofcontents


\phantomsection
\pdfbookmark[1]{Figures list}{sec:figures} 
\listoffigures


\phantomsection
\pdfbookmark[1]{Table list}{sec:tables} 
\listoftables


\phantomsection
\pdfbookmark[1]{Code list}{sec:listings} 
\lstlistoflistings

\pagebreak


\section{Anti-aliasing}
\label{sec:anti-aliasing}

Aliasing occurs when floating-point data of polygon position is decimated to the increments of monitor pixel width and height. Anti-aliasing techniques tend to emulate high-precision data through pixels blending.
This approach prones to be resource-heavy. With techniques like MSAA, SMAA, FXAA suffering from temporal aliasing, where small features and lines is motion tend to flicker.
Here technique like TSAA aims to address this problem, at a cost of performance.
\paragraph[Rasterization Map Anti-Aliasing]{Rasterization Map Anti-Aliasing \textmd{(RMAA)}} presented in this section takes different approach. High-resolution data is already stored in floating-point vertex position. RMAA transfers that precision into pixel's value, effectively decimating at a discrete resolution of $(2^{\textit{bits}}\times\textit{resolution})$. Which is $2^8$-bigger for an 8-bit picture, than the standard rendering. For an 8-bit 1920$\times$1080-HD picture the discrete value is equal 491K ($491\,520\times276\,480$ pixels).
Additionally visible data is always rendered, even at sub-pixel thickness. This reduces temporal artifacts.
The actual real rasterization resolution for RMAA is native to the screen, such that no multi-sampling occurs.

Since resulting coverage mask is non-binary, multiple polygons can cover same pixel at the intersection point, leading to multiple samples per pixel. This is especially present with micro-polygon geometry.

RMAA is not a post-process and does not sample buffer's neighboring pixels.
This makes performance of the algorithm superior to other anti-aliasing techniques. In temporal and visual quality RMAA is comparable to MSAA$\times16+$ for 8-bit picture, and MSAA$\times32+$ for a 10-bit HDR picture.

The anti-aliasing effect is achieved through half-space rasterization,\supercite{Pineda1988EdgeFunctions} with a modified \emph{pixel step} function (firstly introduced in a previous work)\supercite{Fober2020VisualSphere}.
\begin{equation}
\begin{split}
	\text{pixStep}(\Gamma) &= \left\{
		\frac{\Gamma}{|\vec{\nabla\Gamma}|}
		+\frac{1}{2}\right
	\}\cap[0,1]
\\
	&=
	\left\{
		\Gamma\div
		\sqrt{
			\left(\frac{\partial \Gamma}{\partial x}\right)^2+
			\left(\frac{\partial \Gamma}{\partial y}\right)^2
		}+\frac{1}{2}
	\right\}\cap[0,1]
\end{split}
\end{equation}
Here $\Gamma$ represents polygon-edge gradient, with value of zero at the edge, positive value towards polygon center and negative away.
Partial derivatives $\nicefrac{\partial \Gamma}{\partial x}$ and $\nicefrac{\partial \Gamma}{\partial y}$ are equivalent of GLSL's functions $\textbf{dFdx}(\Gamma)$ and $\textbf{dFdy}(\Gamma)$. See listing \vref{lst:anti-aliasing} for more information.

\begin{figure}[p]
	\begin{subfigure}{\textwidth}
		\begin{align}
			\textbit{length}(\textbit{vec2}\big(\textbit{dFdx}\big(\Gamma\big), \textbit{dFdy}\big(\Gamma\big)\big))
			&\equiv
			|\vec{\nabla\Gamma}|
		\tag{\textit{I}}
		& \qquad&\qquad \\
			\textbit{fwidth}(\Gamma)
			&\equiv
			\left|\frac{\partial\Gamma}{\partial x}\right|+
			\left|\frac{\partial\Gamma}{\partial y}\right|
		\tag{\textit{II}}
		& \qquad&\qquad \\
			\textbit{length}(\textbit{vec2}\big(\textbit{dFdx}\big(\Gamma\big), \textbit{dFdy}\big(\Gamma\big)\big)*2)
			&\equiv
			|2\vec{\nabla\Gamma}|
		\tag{\textit{III}}
		& \qquad&\qquad
		\end{align}
		\caption[Edge-slope equations]{Three different edge-slope functions for aliasing-free half-space rasterization.}
		\label{fig:slope equations}
	\end{subfigure}
\\
	\begin{subfigure}{\textwidth}
		\centerline{\includegraphics{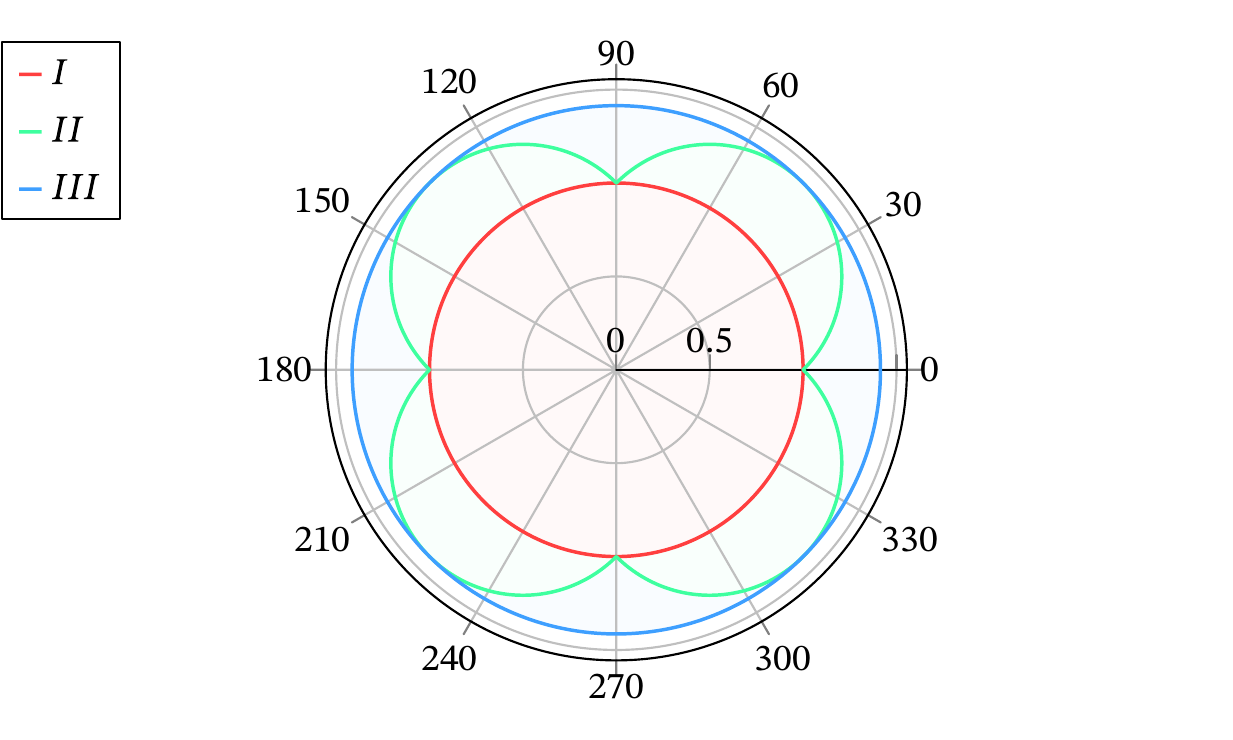}}
		\caption[Slope in radial graph]{Edge slope graph of various step functions, where $r\in\text{pixels}$. Notice how step function $II$ has varying slope $r$, with its maximum diagonally and minimum horizontally \& vertically.}
		\label{fig:slope width graph}
	\end{subfigure}
\\
	\begin{subfigure}{\textwidth}
		\centerline{\includegraphics{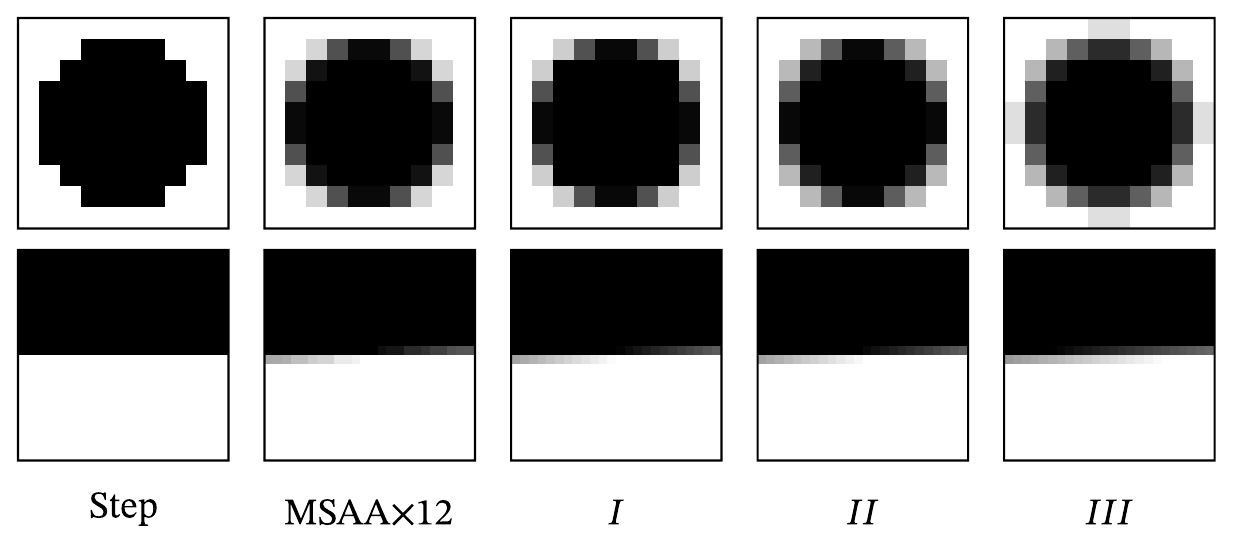}}
		\caption[Rasterized edge and circle]{Edge and circle rasterized with various edge-step functions. Notice how step function $I$ resembles MSAA$\times$12 the most. MSAA$\times$12 also exhibits low tonal-resolution of 144.}
		\label{fig:gradient edge circle}
	\end{subfigure}

	\caption[Step-functions slope]
	{Edge-slope comparison for various step functions.}
	\label{fig:step-functions slope}
\end{figure}

\begin{lstlisting}[float=p,
	label={lst:anti-aliasing},
	caption={
		[Anti-aliasing functions]
		RMAA step functions in GLSL.
	}
]
// Edge rasterization function
float pixStep(float grad)
{
	vec2 Del = vec2(dFdx(grad), dFdy(grad));
	return clamp(inversesqrt(dot(Del, Del))*grad+0.5, 0.0, 1.0);
}
// Triangle edges rasterization function
vec3 pixStep(vec3 gradVec)
{
	mat3x2 DelMtx = transpose( mat2x3(
		dFdx(gradVec), dFdy(gradVec)
	) );
	return clamp(inversesqrt( vec3(
		dot(DelMtx[0], DelMtx[0]),
		dot(DelMtx[1], DelMtx[1]),
		dot(DelMtx[2], DelMtx[2])
	) )*gradVec+0.5, 0.0, 1.0);
}
\end{lstlisting}

\subsection[Rasterization with RMAA]{RMAA rasterization process}
\label{sub:rmaa rasterization}

Rasterization process with RMAA does not differ much from the current half-space rasterization pipeline. It includes edge-function rasterization matrix $\chi\in\real^{3\times3}$ and multiplication of such by fragment's pixel or texture coordinates.
Hard-\emph{step} of the edge-function is replaced by a \emph{pixel-step} function. Special blending procedure is performed when merging fragment data with the buffer (since coverage mask is non-binary).
Additionally bounding box with RMAA technique is expanded by one pixel (half-pixel in each direction), compensating for the edge blur.
\begin{rem}
	Projected vertices that constitute rasterization matrix must be in the same space as $\vec f$ vector (\emph{pixel} or \emph{texture}, \emph{centered} or \emph{cornered}, \emph{square} or \emph{non-square} coordinate).
\end{rem}

\begin{equation}
\begin{split}
	\chi &=
	\begin{bmatrix}
		\vec B_y-\vec C_y &\
		\vec C_x-\vec B_x &\
		\vec B_x\vec C_y-\vec B_y\vec C_x
		\smallskip\\
		\vec C_y-\vec A_y &\
		\vec A_x-\vec C_x &\
		\vec C_x\vec A_y-\vec C_y\vec A_x
		\smallskip\\
		\vec A_y-\vec B_y &\
		\vec B_x-\vec A_x &\
		\vec A_x\vec B_y-\vec A_y\vec B_x
	\end{bmatrix} \qed
\\
	&=
	\begin{bmatrix}
		\Big[\vec B_x & \vec B_y & 1\Big]
		&\cross&
		\Big[\vec C_x & \vec C_y & 1\Big]
	\smallskip\\
		\Big[\vec C_x & \vec C_y & 1\Big]
		&\cross&
		\Big[\vec A_x & \vec A_y & 1\Big]
	\smallskip\\
		\Big[\vec A_x & \vec A_y & 1\Big]
		&\cross&
		\Big[\vec B_x & \vec B_y & 1\Big]
	\end{bmatrix}
	\qquad
	\begin{matrix}
		\text{edge }f\big(a\big) \\[0.62em]
		\text{edge }f\big(b\big) \\[0.62em]
		\text{edge }f\big(c\big)
	\end{matrix}
\end{split}
\end{equation}
$\chi\in\real^{3\times3}$ is the $\overline{ABC}$'s triangle rasterization matrix, for a counter-clock-wise polygon. The matrix is calculated once-per visible polygon, and sent as a rasterizer input. Triangle points $\{\vec A,\vec B,\vec C\}\in\real^2$ represent vertex positions in texture or pixel coordinates, after perspective division. Each row vector of matrix $\chi$ is a triangle-edge function ($a$, $b$, $c$ respectively). Lines are labeled after the opposite vertex.
Fist two columns of the matrix $\chi$ represent scaled $\sin\alpha$ and $\cos\alpha$ of the rotation angle $\alpha$, while third -- the line offset. See listing \vref{lst:rectilinear rasterization} for more information.

Each row-vector of rasterization matrix $\chi$ is derived from equation like the following:
\begin{equation}
\begin{split}
	\begin{bmatrix}
		\vec f_x-\vec A_x \\
		\vec f_y-\vec A_y
	\end{bmatrix}
	\cdot
	\begin{bmatrix}
		-(\vec B_y-\vec A_y) \\
		\vec B_x-\vec A_x
	\end{bmatrix}
	&=
	\big(\vec f_x-\vec A_x\big)\big(\vec A_y-\vec B_y\big)+
	\big(\vec f_y-\vec A_y\big)\big(\vec B_x-\vec A_x\big)
	\\&=
	\vec f_x\big(\vec A_y-\vec B_y\big)+
	\vec f_y\big(\vec B_x-\vec A_x\big)+
	\vec A_x\vec B_y-\vec A_y\vec B_x
	\\&=
	\begin{bmatrix}
		\vec f_x \\ \vec f_y \\ 1
	\end{bmatrix}
	\cdot
	\begin{bmatrix}
		A_y-B_y \\ B_x-A_x \\ A_xB_y-A_yB_x
	\end{bmatrix}
\end{split}
\end{equation}
Result of above equation is a gradient, with value of zero at $\overline{AB}$ line-edge, above zero towards $\vec C$ and negative away from $\vec C$ (for counter-clock-wise triangle).
Combination of three such equations (for corresponding triangle edges) form a half-space triangle $\vec\Xi$.

\begin{subequations}
\begin{align}
	\begin{bmatrix}
		\vec \Xi_1 \\ \vec \Xi_2 \\ \vec \Xi_3
	\end{bmatrix}
	&=
	\begin{bmatrix}
		\vec f_x \\ \vec f_y \\ 1
	\end{bmatrix}
	\begin{bmatrix}
		\chi_{1,1} & \chi_{1,2} & \chi_{1,3} \\
		\chi_{2,1} & \chi_{2,2} & \chi_{2,3} \\
		\chi_{3,1} & \chi_{3,2} & \chi_{3,3}
	\end{bmatrix}
	\\
	\Lambda &= \prod^3_{i=1}\text{pixStep}\big(\Xi_i\big) \qed
\end{align}
\end{subequations}
Aliasing-free polygon coverage mask $\Lambda\in[0,1]$ is a product of triangle half-space components $\vec\Xi\in\real^3$ after the pixel-step function. See listing \vref{lst:rectilinear rasterization} for more information.
\begin{rem}
	RMAA technique requires \emph{gamma}-correction of the final composite picture from \emph{linear-color-space} in order to achieve desired smooth result.
\end{rem}

\paragraph[Bounding box]{Bounding box} width and height is expanded by one pixel. This compensates for the blur-width of anti-aliased edge.

\begin{equation}
	\begin{bmatrix}
		B_{1,1} & B_{1,2} \\
		B_{2,1} & B_{2,2}
	\end{bmatrix}
	=
	\begin{bmatrix}
		\left\lceil
		\min\left\{
			\vec A_s,
			\vec B_s,
			\vec C_s\right\}-\nicefrac{1}{2}
		\right\rceil
	&
		\left\lceil
		\min\left\{
			\vec A_t,
			\vec B_t,
			\vec C_t\right\}-\nicefrac{1}{2}
		\right\rceil
	\\[1em]
		\left\lfloor
		\max\left\{
			\vec A_s,
			\vec B_s,
			\vec C_s\right\}+\nicefrac{1}{2}
		\right\rfloor
	&
		\left\lfloor
		\max\left\{
			\vec A_t,
			\vec B_t,
			\vec C_t\right\}+\nicefrac{1}{2}
		\right\rfloor
	\end{bmatrix}
\end{equation}
Bounding box coordinate matrix $B\in\positive_0^{2\times2}$ expanded by half-pixel in each direction. First row represents bottom-left corner viewport-position and second, top-right corner. See listing \vref{lst:rectilinear rasterization} for more information.

\paragraph{Rectilinear rasterization} has constant partial derivative $\partial\Gamma$, equal to one pixel. Therefore transformation inside $\text{pixStep}(\Gamma)$ function can be integrated into rasterization matrix $\chi\in\real^{3\times3}$.
Using viewport triangle coordinates $\{\vec A,\vec B,\vec C\}\in\real^2_{>0}$ with floating-point precision (sub-pixel position).

\begin{equation}
	\chi' =
	\begin{bmatrix}
		\frac{\chi_{1,1}}{|[\chi_{1,1} \quad \chi_{1,2}]|} &\
		\frac{\chi_{1,2}}{|[\chi_{1,1} \quad \chi_{1,2}]|} &\
		\frac{\chi_{1,3}}{|[\chi_{1,1} \quad \chi_{1,2}]|}+\frac{1}{2}
	\smallskip\\
		\frac{\chi_{2,1}}{|[\chi_{2,1} \quad \chi_{2,2}]|} &\
		\frac{\chi_{2,2}}{|[\chi_{2,1} \quad \chi_{2,2}]|} &\
		\frac{\chi_{2,3}}{|[\chi_{2,1} \quad \chi_{2,2}]|}+\frac{1}{2}
	\smallskip\\
		\frac{\chi_{3,1}}{|[\chi_{3,1} \quad \chi_{3,2}]|} &\
		\frac{\chi_{3,2}}{|[\chi_{3,1} \quad \chi_{3,2}]|} &\
		\frac{\chi_{3,3}}{|[\chi_{3,1} \quad \chi_{3,2}]|}+\frac{1}{2}
	\end{bmatrix}
	\qquad
	\begin{matrix}
		\text{edge }f\big(a\big) \\[0.86em]
		\text{edge }f\big(b\big) \\[0.86em]
		\text{edge }f\big(c\big)
	\end{matrix}
\end{equation}
Rasterization matrix $\chi'$ for counter-clock-wise, viewport-space triangle $\overline{ABC}$. Row 2D-normalization preserves gradient slope. Additional $\nicefrac{1}{2}$ pixel offset sets gradient center at the edge. See listing \vref{lst:rectilinear rasterization} for more information.
\begin{rem}
	Perspective rasterization matrix $\chi\in\real^{3\times3}$ can be produced from vertex data before perspective division, using \emph{homogeneous clip-space} points $\{\vec A,\vec B,\vec C\}\in\real^3$ and their \emph{cross product}. But it requires centered and aspect-correct coordinates at the rasterization-stage.
\end{rem}

\begin{subequations}
\begin{align}
	\begin{bmatrix}
		\vec \Xi_1 \smallskip\\
		\vec \Xi_2 \smallskip\\
		\vec \Xi_3
	\end{bmatrix}
	&=
	\left\{
	\begin{bmatrix}
		\vec f^p_s \\
		\vec f^p_t \smallskip\\
		1
	\end{bmatrix}
	\begin{bmatrix}
		\chi'_{1,1} & \chi'_{1,2} & \chi'_{1,3} \smallskip\\
		\chi'_{2,1} & \chi'_{2,2} & \chi'_{2,3} \smallskip\\
		\chi'_{3,1} & \chi'_{3,2} & \chi'_{3,3}
	\end{bmatrix}
	\right\} \cap [0,1]
	\smallskip\\
	\Lambda &=
	\prod^3_{i=1} \vec \Xi_i \qed
\end{align}
\end{subequations}
Rasterization of aliasing-free polygon coverage-mask $\Lambda\in[0,1]$, here $\vec f^p\in\positive_0^2$ denotes fragment's pixel-index coordinate. Coverage mask $\Lambda$ is formed as a product of clamped $\vec\Xi\in[0,1]^3$ components, where each represents a half-space. See listing \vref{lst:rectilinear rasterization} for more information.
\begin{rem}
	Since rectilinear rasterization does not utilize $\textbit{dFdx}(\Gamma)$ or $\textbit{dFdy}(\Gamma)$ function, RMAA can be successfully executed with software rasterization.
\end{rem}

\begin{lstlisting}[float=p,
	label={lst:rectilinear rasterization},
	caption={
		[Rectilinear rasterization]
		Rectilinear rasterization functions in GLSL.
	}
]
// Get rectilinear edge function
vec3 getEdgeFunc(vec2 P0, vec2 P1)
{
	vec3 edgeFunc = vec3( P0.y-P1.y, P1.x-P0.x,
		P0.x*P1.y-(P0.y*P1.x) );
	// 2D normalization for gradient-slope preservation
	return edgeFunc*inversesqrt(dot(edgeFunc.xy, edgeFunc.xy));
}
// Get rectilinear rasterization matrix
mat3 getEdgeMtx(mat3 polyg)
{
	return mat3(
		getEdgeFunc(polyg[1].xy, polyg[2].xy), // Line a
		getEdgeFunc(polyg[2].xy, polyg[0].xy), // Line b
		getEdgeFunc(polyg[0].xy, polyg[1].xy)  // Line c
	);
}
// Edge function evaluation
float fedge(vec2 pos, vec3 edge)
{ return pos.x*edge.s+(pos.y*edge.t+edge.p); }
// Edge function slope to barycentric (weight)
vec3 edgeToBarycentric(mat3 rastMtx, mat3 polyg)
{
	return 1.0/vec3(
		fedge(polyg[0].xy, rastMtx[0]),
		fedge(polyg[1].xy, rastMtx[1]),
		fedge(polyg[2].xy, rastMtx[2]) );
}
// Returns depth and perspective-correct barycentric coordinates
float correctBarycentric(inout vec3 barycentric, vec3 rcpVrtxDepth)
{
	float depth = 1.0/dot(rcpVrtxDepth, barycentric);
	barycentric *= depth*rcpVrtxDepth; return depth; // Output
}
// Get bounding box coordinates
mat2 getBB(mat3 polyg)
{
	return mat2(
		min(min(polyg[0].xy, polyg[1].xy), polyg[2].xy)-0.5,
		max(max(polyg[0].xy, polyg[1].xy), polyg[2].xy)+0.5 );
}
// Rasterize triangle mask with RMAA
vec4 rasterize(ivec2 pixcoords, mat3 rasterMtx, vec3 barycWeight)
{
	vec3 halfSpace = vec3(
		fedge(pixcoords, rasterMtx[0]),
		fedge(pixcoords, rasterMtx[1]),
		fedge(pixcoords, rasterMtx[2]) );
	vec3 barycentric = halfSpace*barycWeight;
	halfSpace = clamp(halfSpace+0.5, 0.0, 1.0);
	// Return barycentric coordinates and coverage mask
	return vec4(barycentric, halfSpace.r*halfSpace.g*halfSpace.b);
}
\end{lstlisting}

\paragraph{Barycentric coordinate}  interpolates vertex data across polygon-surface.
It is possible to obtain barycentric vector in linear form using scaled edge function matrix $\chi$.

\begin{align}
	\chi &=
	\begin{bmatrix}
		\left[\frac{\vec B\cross\vec C}{\vec A\cdot(\vec B\cross\vec C)}\right]\transpose \smallskip\\
		\left[\frac{\vec C\cross\vec A}{\vec B\cdot(\vec C\cross\vec A)}\right]\transpose \smallskip\\
		\left[\frac{\vec A\cross\vec B}{\vec C\cdot(\vec A\cross\vec B)}\right]\transpose
	\end{bmatrix}
	\qquad
	\begin{matrix}
		\text{edge }f\big(a\big) \\[1.16em]
		\text{edge }f\big(b\big) \\[1.16em]
		\text{edge }f\big(c\big)
	\end{matrix}
	\label{eq:barycentric matrix}
\\
	\begin{bmatrix}
		\vec\lambda_s \\ \vec\lambda_t \\ \vec\lambda_p
	\end{bmatrix}
	&=
	\begin{bmatrix}
		\vec f_x \\ \vec f_y \\ 1
	\end{bmatrix}
	\begin{bmatrix}
		\chi_{1,1} & \chi_{1,2} & \chi_{1,3} \\
		\chi_{2,1} & \chi_{2,2} & \chi_{2,3} \\
		\chi_{3,1} & \chi_{3,2} & \chi_{3,3}
	\end{bmatrix}
\end{align}
Edge-function matrix $\chi$ doubles as a barycentric coordinate transformation matrix.
Points $\{\vec A,\vec B,\vec C\}$ represent vertex position after perspective division. $\vec\lambda\in\real^3$ is the linear-barycentric coordinate, while $\vec f\in\positive^2_0\text{ or }[0,1]^2$ denotes pixel or texture frame coordinates. See listing \vref{lst:rectilinear rasterization} for more information.

It is possible to separate denominator from barycentric transformation matrix $\chi$ and use it as a barycentric-edge weight $\vec\omega\in\real^3_{>0}$. Calculated per triangle, $\vec\omega$ can convert triangle's unclipped half-space gradient into linear-barycentric vector.

\begin{subequations}
\begin{align}
	\begin{bmatrix}
		\vec\omega_s \\
		\vec\omega_t \\
		\vec\omega_p
	\end{bmatrix}
	&=
	\begin{bmatrix}
		\vec A\cdot\begin{bmatrix}\chi_{1,1} & \chi_{1,2} & \chi_{1,3}\end{bmatrix} \\
		\vec B\cdot\begin{bmatrix}\chi_{2,1} & \chi_{2,2} & \chi_{2,3}\end{bmatrix} \\
		\vec C\cdot\begin{bmatrix}\chi_{3,1} & \chi_{3,2} & \chi_{3,3}\end{bmatrix}
	\end{bmatrix}^{-1}
\\
	\begin{bmatrix}
		\vec\Xi_1 \\
		\vec\Xi_2 \\
		\vec\Xi_3
	\end{bmatrix}
	&=
	\begin{bmatrix}
		\vec f^p_s \\
		\vec f^p_t \smallskip\\
		1
	\end{bmatrix}
	\begin{bmatrix}
		\chi'_{1,1} & \chi'_{1,2} & \chi'_{1,3} \smallskip\\
		\chi'_{2,1} & \chi'_{2,2} & \chi'_{2,3} \smallskip\\
		\chi'_{3,1} & \chi'_{3,2} & \chi'_{3,3}
	\end{bmatrix}
	\tag{half-space}
\\
	\Lambda &=
	\prod^3_{i=1} \left\{ \vec \Xi_i+\nicefrac{1}{2} \right\} \cap [0,1] \qed
	\tag{coverage}
\\
	\begin{bmatrix}
		\vec\lambda_s \\
		\vec\lambda_t \\
		\vec\lambda_p
	\end{bmatrix}
	&=
	\begin{bmatrix}
		\vec\Xi_1 \\
		\vec\Xi_2 \\
		\vec\Xi_3
	\end{bmatrix}
	\begin{bmatrix}
		\omega_s \\
		\omega_t \\
		\omega_p
	\end{bmatrix} \qed
\end{align}
\end{subequations}
Here $\vec\omega\in\real^3_{>0}$ represents barycentric edge-weight vector. It converts slope of the edge function into linear-barycentric coordinate slope. $\vec\Xi\in\real^3$ denotes half-space triangle, and $\Lambda\in[0,1]$ its coverage mask. See listing \vref{lst:rectilinear rasterization} for more information.

\paragraph{Depth pass} of the fragment $\vec f_z$ is obtained using inverse of the dot product between barycentric vector $\vec\lambda$ and inverse vertex-depth $\big\{\vec A_z^{-1},\vec B_z^{-1},\vec C_z^{-1}\big\}$ vector. This is a standard graphics-pipeline procedure.

\begin{subequations}
\begin{align}
	\vec f_z &=
	\left(
	\begin{bmatrix}
		\vec \lambda_s \\
		\vec \lambda_t \\
		\vec \lambda_p
	\end{bmatrix}
	\cdot
	\begin{bmatrix}
		\vec A^{-1}_z \\
		\vec B^{-1}_z \\
		\vec C^{-1}_z
	\end{bmatrix}
	\right)^{-1}
\\
	\begin{bmatrix}
		\vec \lambda'_s \\
		\vec \lambda'_t \\
		\vec \lambda'_p
	\end{bmatrix}
	&=
	\vec f_z
	\begin{bmatrix}
		\vec \lambda_s \\
		\vec \lambda_t \\
		\vec \lambda_p
	\end{bmatrix}
	\begin{bmatrix}
		\vec A^{-1}_z \\
		\vec B^{-1}_z \\
		\vec C^{-1}_z
	\end{bmatrix}
	\label{eq:barycentric correct}
\end{align}
\end{subequations}
Here $\vec f_z\in\real_{>0}$ denotes pixel-depth value, where $\vec\lambda'\in\real^3$ represent perspective-correct barycentric vector. $\vec\lambda'$ is used for interpolation of vertex-data across polygon surface. See listing \vref{lst:rectilinear rasterization} for more information.

\subsection{Fragment merging procedure}
\label{sub:fragment merging}

Since aliasing-free rasterization coverage mask is non-binary, buffer merging procedure is required. This procedure works with front-to-back rasterization. Such ordered rasterization procedure could be achieved through binary space partitioning (BSP).\supercite{Dupuy2020ConcurrentBinaryTrees,Fuchs1980BinarySpacePartitioning,Newell1972hidden_surface} It also supports anti-aliased alpha-to-coverage, through texture-filtering.

\begin{align}
	\Lambda^{f\prime}
	&=
	\begin{cases}
		\min\{\Lambda^f,1-\Lambda^b\}\cdot \Lambda_a, & \text{if \emph{alpha-blending}} \\
		\min\{\Lambda^f,1-\Lambda^b\}, & \text{otherwise}
	\end{cases}
	\tag{mask clipping}
\\
	\Lambda^{b\prime}
	&=
	\Lambda^b+\Lambda^{f\prime}
	\tag{mask merging}
\end{align}
$\Lambda^f\in[0,1]$ denotes current fragment mask, $\Lambda^b\in[0,1]$ is the geometry-mask buffer with $\Lambda_a\in[0,1]$ as the fragment-alpha texture mask. Therefore $\Lambda^{f\prime}\in[0,1]$ represents fragment mask clipped by occluding geometry of the buffer.
Finally $\Lambda^{b\prime}\in[0,1]$ is the merged coverage buffer.
This merging and clipping procedure requires ordered rasterization (in this case front--to--back).
See listing \vref{lst:buffering} for more information.

Same mixing principle can be applied for adding fragment pass to the buffer. For example in deferred shading, normal pass $\hat N$, world position pass $\vec W$, diffuse texture and PBR textures can all use same buffer merging procedure.
\begin{gather}
	F^b(1) = 0
	\tag{initial buffer}
\\
	F^{b\prime} = \overbrace{\Lambda^{f\prime}F^f+F^b}^\text{front-to-back}
	\tag{data pass}
\end{gather}
$F^b$ denotes buffered data, with $F^f$ as the current fragment data. Finally $F^{b\prime}$ represents merged buffer data. See listing \vref{lst:buffering} for more information.
\begin{rem}
	With aliased rasterization, fragment-clipping and buffer-merging can be performed with simple (binary) depth-test, without order of rasterization.
\end{rem}

\begin{lstlisting}[float=p,
	label={lst:buffering},
	caption={
		[Blending fragment with buffer]
		RMAA-rasterization fragment data blending procedures, in GLSL.
	}
]
// Clip fragment mask in front-back rasterizer
float clipMsk(float fragMsk, float buffMsk)
{ return min(fragMsk, 1.0-buffMsk); }
float clipMsk(float fragMsk, float fragAlpha, float buffMsk)
{ return min(fragMsk, 1.0-buffMsk)*fragAlpha; }

// Blend float buffer pass in front-back rasterizer
float bufferMerge(float fragClpMsk, float fragData, float buffData)
{ return fragClpMsk*fragData+buffData; }
// Blend vec2 buffer pass in front-back rasterizer
vec2 bufferMerge(float fragClpMsk, vec2 fragData, vec2 buffData)
{ return fragClpMsk*fragData+buffData; }
// Blend vec3 buffer pass in front-back rasterizer
vec3 bufferMerge(float fragClpMsk, vec3 fragData, vec3 buffData)
{ return fragClpMsk*fragData+buffData; }
// Blend vec4 buffer pass in front-back rasterizer
vec4 bufferMerge(float fragClpMsk, vec4 fragData, vec4 buffData)
{ return fragClpMsk*fragData+buffData; }
\end{lstlisting}

\section[Rasterization map]{Rasterization map}
\label{sec:rasterization map}

Since rasterization process of RMAA technique utilizes only screen coordinates for geometry drawing, those coordinates can be encoded in a lookup texture. For example commonly known \emph{STMap} format could be used.
There are two variants for each coordinate-lookup texture, distorting and undistorting.
\begin{description}
	\item[Distorting]
		\emph{STMap} stores texture coordinates $\vec f$ in distorted space $\vec G$, effectively mapping between $\vec G_{st}\mapsto\vec f_{st}$. This texture map is used for rasterization with distortion and for sampling image through distorted space.
		User screen-space input can be mapped to world space using distortion \emph{STMap}.
	\item[Undistorting]
		\emph{STMap} stores coordinates of distorted space $\vec G$ in undistorted $\vec f$ texture coordinates, effectively mapping between $\vec f_{st}\mapsto\vec G_{st}$. This texture map is used for GUI positioning (from world space, to screen space). It is also used to undistort live footage for motion tracking.
		\begin{rem}
			To note, some non-linear projections cannot be satisfied with undistort \emph{STMap}, since single world-space point can occupy multiple screen-space positions in some projections.
		\end{rem}
		\begin{example}
			In equidistant fish-eye projection at $\Omega=360\degree$, point opposite to the view-direction forms a ring around the picture border.
		\end{example}
		\begin{example}
			In equirectangular projection, north and south pole forms two horizontal lines.
		\end{example}
\end{description}
\begin{figure}[h]
	\centering
	\begin{subfigure}{140pt}
		\centerline{\includegraphics{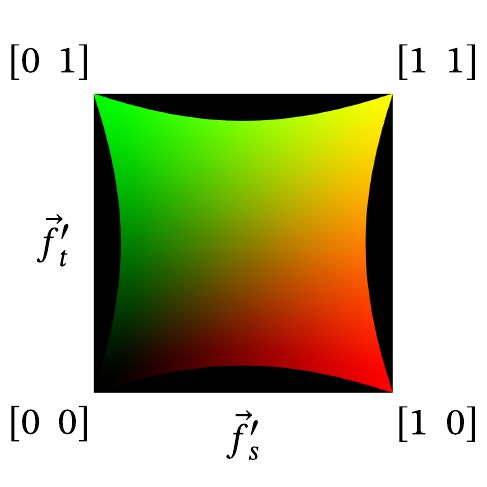}}
		\captionsetup{justification=centering}
		\caption[Undistort STMap]{Undistort \emph{STMap} with \\$\Omega^d=90\degree$, $k=-1$.}
		\label{fig:undistort STMap}
	\end{subfigure}
	\hspace{1.382em}
	\begin{subfigure}{140pt}
		\centerline{\includegraphics{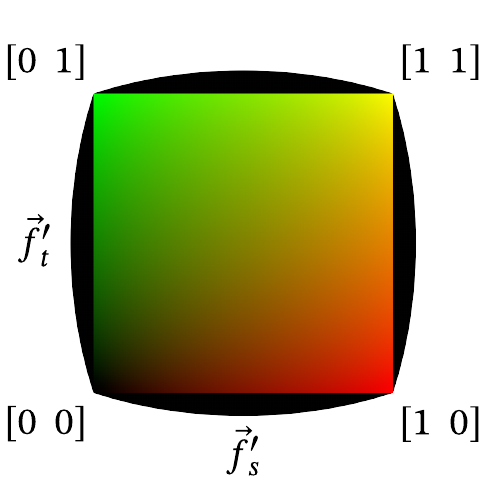}}
		\captionsetup{justification=centering}
		\caption[Distort STMap]{Distort \emph{STMap} with \\$\Omega^d=90\degree$, $k=-1$.}
		\label{fig:distort STMap}
	\end{subfigure}
	\caption[Distort and undistort STMaps]{Example of distort and undistort \emph{STMap}.
	\emph{Red} and \emph{Green} are the \textquote*{$st$} coordinate components of $\vec f'\in[0,1]^2$.
	Black represents out-of-bounds pixels.}
	\label{fig:distort undistort STMaps}
\end{figure}

\subsection[Rasterization with STMap]
{Rasterization with lookup texture coordinates}
\label{sub:lookup rasterization}

Rasterization process with \emph{STMap} differs from the rectilinear rasterization.
Since partial derivative $\partial\Gamma$ is not constant with distorted \emph{STMap}, it has to be calculated at the rasterization step. On the other hand, rasterization matrix $\chi$ can be modified to output barycentric coordinates in linear-variant, in a same way as the equation \vref{eq:barycentric matrix}. Result barycentric vector doubles then as half-space triangle gradient.

\begin{equation}
	\begin{bmatrix}
		\chi_{1,1} & \chi_{1,2} & \chi_{1,3} \\
		\chi_{2,1} & \chi_{2,2} & \chi_{2,3} \\
		\chi_{3,1} & \chi_{3,2} & \chi_{3,3}
	\end{bmatrix}
	=
	\begin{bmatrix}
		\bigg[\frac{\vec B\cross\vec C}{\vec A\cdot(\vec B\cross\vec C)}\bigg]\transpose 
	\\[0.62em]
		\bigg[\frac{\vec C\cross\vec A}{\vec B\cdot(\vec C\cross\vec A)}\bigg]\transpose 
	\\[0.62em]
		\bigg[\frac{\vec A\cross\vec B}{\vec C\cdot(\vec A\cross\vec B)}\bigg]\transpose 
	\end{bmatrix}
	\qquad
	\begin{matrix}
		\text{edge }f(a) \\[1.38em]
		\text{edge }f(b) \\[1.38em]
		\text{edge }f(c)
	\end{matrix}
\end{equation}
Rasterization matrix $\chi$ for counter-clock-wise triangle, with linear-barycentric vector as an output. It is calculated using $\{\vec A, \vec B, \vec C\}\in[0,1]^3$ vertex coordinates in \emph{STMap} space, after perspective division (such that $z=1$). Division by the dot-product neglects winding of the triangle, making back-facing polygons visible. For more information see listing \vref{lst:stmap rasterization}.

\begin{subequations}
\begin{align}
	\begin{bmatrix}
		\vec\lambda_s \\
		\vec\lambda_t \\
		\vec\lambda_p
	\end{bmatrix}
	&=
	\begin{bmatrix}
		\vec f'_s \\
		\vec f'_t \\
		1
	\end{bmatrix}
	\begin{bmatrix}
		\chi_{1,1} & \chi_{1,2} & \chi_{1,3} \\
		\chi_{2,1} & \chi_{2,2} & \chi_{2,3} \\
		\chi_{3,1} & \chi_{3,2} & \chi_{3,3}
	\end{bmatrix}
\\
	\begin{bmatrix}
		\vec\Xi_1 \smallskip\\
		\vec\Xi_2 \smallskip\\
		\vec\Xi_3
	\end{bmatrix}
	&=
	\left\{
	\vphantom{\begin{bmatrix}
		\vec\lambda_s \smallskip\\
		\vec\lambda_t \smallskip\\
		\vec\lambda_p
	\end{bmatrix}}
	\right.
	\overbrace{\frac{1}{2}+
	\begin{bmatrix}
		{|\vec{\nabla\lambda_s}|}^{-1}
	\smallskip\\
		{|\vec{\nabla\lambda_t}|}^{-1}
	\smallskip\\
		{|\vec{\nabla\lambda_p}|}^{-1}
	\end{bmatrix}}^\text{anti-aliasing}
	\left.
	\begin{bmatrix}
		\vec\lambda_s \smallskip\\
		\vec\lambda_t \smallskip\\
		\vec\lambda_p
	\end{bmatrix}
	\right\}
	\cap [0,1]
\\
	\Lambda &= \prod^3_{i=1} \vec \Xi_i \qed
\end{align}
\end{subequations}
Above is two stage rasterization process, where $\vec f'$ represents the \emph{STMap} coordinate.
At first, edge function matrix $\chi$ multiplication establishes linear-barycentric coordinate $\vec\lambda\in\real^3$. Secondly partial derivative-inverse scales the edge slope to one pixel width, creating aliasing-free half-space triangle $\vec\Xi\in[0,1]^3$. Later triangle coverage mask $\Lambda\in[0,1]$ is formed as a product of $\vec\Xi$'s components.
For more information on that see listing \vref{lst:stmap rasterization}.

\begin{lstlisting}[float=p,
	label={lst:stmap rasterization},
	caption={
		[STMap rasterization]
		\emph{STMap} rasterization functions in GLSL.
		For the $\textbit{pixStep}\big(\vec\lambda\big)$ function definition see listing \vref{lst:anti-aliasing}.
	}
]
#define RMAA true

// Get barycentric-edge function
vec3 getEdgeFunc(vec2 P0, vec2 P1, vec2 P2)
{
	vec3 edgeFunc = vec3(
		P0.y-P1.y,
		P1.x-P0.x,
		P0.x*P1.y-(P0.y*P1.x)
	);
	// Convert to barycentric function
	return edgeFunc/(P2.x*edgeFunc.s+(P2.y*edgeFunc.t+edgeFunc.p));
}
// Get barycentric rasterization matrix
mat3 getEdgeMtx(mat3 polyg)
{
	return mat3(
		getEdgeFunc(polyg[1].xy, polyg[2].xy, polyg[0].xy), // Line a
		getEdgeFunc(polyg[2].xy, polyg[0].xy, polyg[1].xy), // Line b
		getEdgeFunc(polyg[0].xy, polyg[1].xy, polyg[2].xy)  // Line c
	);
}
// Edge function evaluation
float fedge(vec2 pos, vec3 edge)
{ return pos.x*edge.s+(pos.y*edge.t+edge.p); }
// Get vector of inverse depth of each triangle point
vec3 getRcpDepth(mat3 polyg)
{ return 1.0/transpose(polyg)[2]; }
// Returns depth and perspective-correct barycentric coordinates
float correctBarycentric(inout vec3 barycentric, vec3 rcpVrtxDepth)
{
	float depth = 1.0/dot(rcpVrtxDepth, barycentric);
	barycentric *= depth*rcpVrtxDepth; return depth; // Output
}
// Rasterize triangle coverage and linear-barycentric coordinates
vec4 rasterize(vec2 STcoords, mat3 rasterMtx)
{
	vec3 barycentric = vec3(
		fedge(STcoords, rasterMtx[0]),
		fedge(STcoords, rasterMtx[1]),
		fedge(STcoords, rasterMtx[2]) );
	vec3 halfSpace = RMAA? pixStep(barycentric) : step(0.0, barycentric);
	// Return barycentric coordinates and coverage mask
	return vec4(barycentric, halfSpace.r*halfSpace.g*halfSpace.b);
}
\end{lstlisting}

\paragraph[Bounding box]{Bounding box \textmd{(BB)}} evaluation is the most challenging process with \emph{STMap} rasterization. It is possible to sample 2D position of BB from undistort \emph{STMap}. Another approach is to test every pixel in the distort \emph{STMap} against BB position.
This can also be performed using MIP-mapped \emph{STMap}, for selective, pyramidal progression to final raster-resolution.
Process can also be achieved through \emph{divide and conquer} technique, where screen is progressively divided into regions of interest.
\begin{rem}
	MIP-mapping with \emph{divide and conquer} technique does not form a bounding box, but rather a render region.
\end{rem}

\begin{subequations}
\begin{align}
	\begin{bmatrix}
		B_{1,1} & B_{1,2} \\
		B_{2,1} & B_{2,2}
	\end{bmatrix}
	&=
	\begin{bmatrix}
		\min\left\{
			\vec A_s,
			\vec B_s,
			\vec C_s\right\} &
		\min\left\{
			\vec A_t,
			\vec B_t,
			\vec C_t\right\}
	\smallskip\\
		\max\left\{
			\vec A_s,
			\vec B_s,
			\vec C_s\right\} &
		\max\left\{
			\vec A_t,
			\vec B_t,
			\vec C_t\right\}
	\end{bmatrix}
\\
	\text{testBB}\big(\vec f'_{st}; B\big) &= \land
	\left\{\begin{aligned}
		\vec f'_s-\nicefrac{1}{2}|\vec{\nabla f'_s}|
		&\geq B_{11}
	\\
		\vec f'_t-\nicefrac{1}{2}|\vec{\nabla f'_t}|
		&\geq B_{12}
	\\
		\vec f'_s+\nicefrac{1}{2}|\vec{\nabla f'_s}|
		&< B_{21}
	\\
		\vec f'_t+\nicefrac{1}{2}|\vec{\nabla f'_t}|
		&< B_{22}
	\end{aligned}\right.
\end{align}
\end{subequations}
Here $\vec f$ represents \emph{STMap} coordinates. First row of the bounding-box matrix $B\in\real^{2\times2}_{\geq0}$ denotes its left-bottom corner, while second row, top-right.

Expression $\vec f_i\pm\nicefrac{1}{2}|\vec{\nabla f_i}|$ can be transformed into MIP-mapped $\vec f''\in\real^4$ texture lookup:

\begin{equation}
	\begin{bmatrix}
		\vec f''_1 \smallskip\\
		\vec f''_2 \smallskip\\
		\vec f''_3 \smallskip\\
		\vec f''_4
	\end{bmatrix}
	=
	\overbrace{
	\begin{bmatrix}
		\vec f'_s \smallskip\\
		\vec f'_t \smallskip\\
		\vec f'_s \smallskip\\
		\vec f'_t
	\end{bmatrix}}^\text{initial MIP-mapped data}
	+
	\underbrace{
	\begin{bmatrix}
		-\nicefrac{1}{2}|\vec{\nabla f'_s}| \smallskip\\
		-\nicefrac{1}{2}|\vec{\nabla f'_t}| \smallskip\\
		 \nicefrac{1}{2}|\vec{\nabla f'_s}| \smallskip\\
		 \nicefrac{1}{2}|\vec{\nabla f'_t}|
	\end{bmatrix}}_\text{transformation per MIP-level}
\end{equation}

\subsection[Universal Perspective STMap]{Universal Perspective \emph{STMap}}
\label{sub:universal perspective}

With ability to render distorted views, comes demand for standardized method of distortion,
not only a corrective one.
A method able to provide image geometry specialized to the desired contend.
Universal Perspective \emph{STMap} model provides the ability to generate distortion of any azimuthal projection with additional choice between cylindrical and spherical projection type.

\begin{figure}[H]
	\centering
	\begin{subfigure}[t]{100pt}
		\includegraphics[width=100pt]{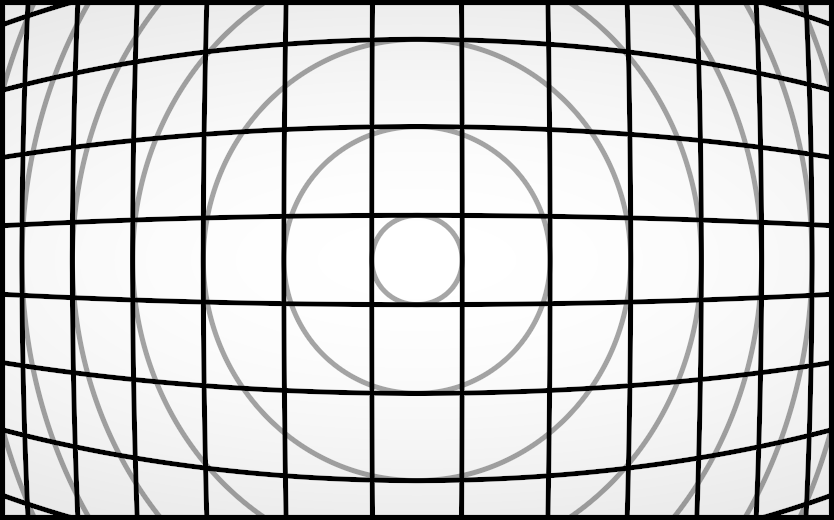}
		\captionsetup{justification=centering}
		\caption{$\Omega^h=90\degree$, $k=-0.5$, $l=0.2$, $s=98\%$}
	\end{subfigure}
	\hspace{16pt}
	\begin{subfigure}[t]{100pt}
		\includegraphics[width=100pt]{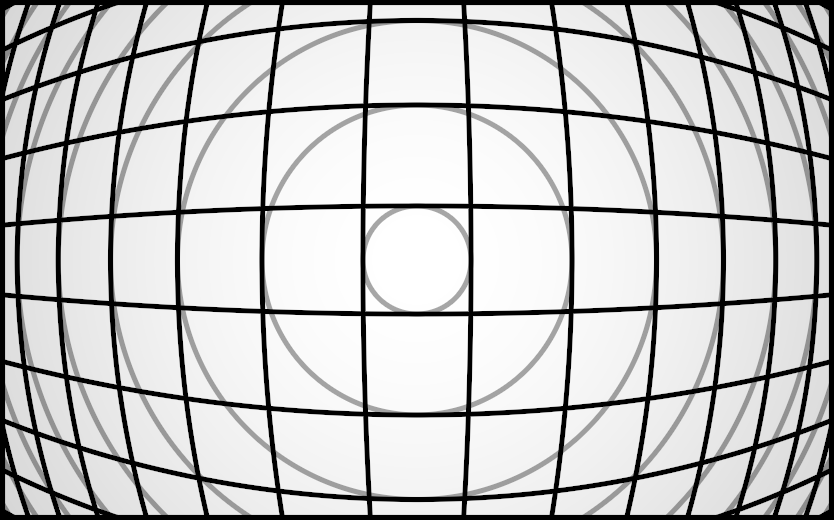}
		\captionsetup{justification=centering}
		\caption{$\Omega^h=124\degree$, $k=0.5$ (stereographic)}
	\end{subfigure}
	\hspace{16pt}
	\begin{subfigure}[t]{100pt}
		\includegraphics[width=100pt]{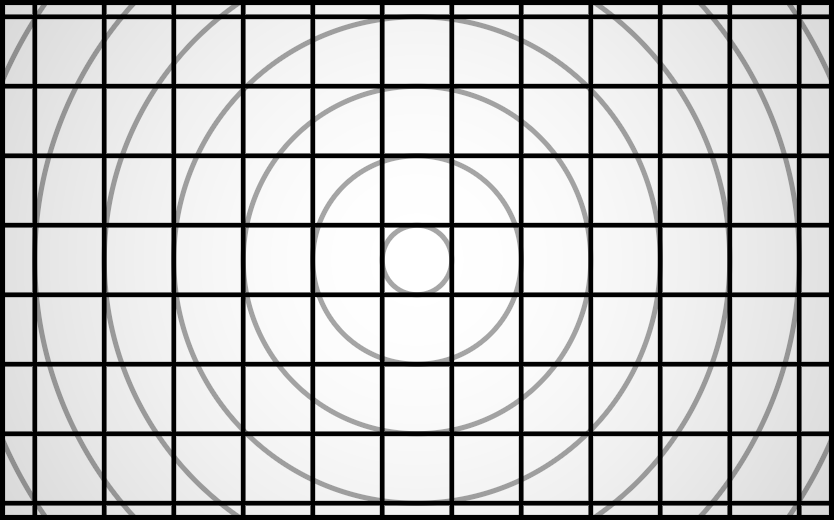}
		\captionsetup{justification=centering}
		\caption{$\Omega^h=60\degree$, $k=1$ (rectilinear)}
	\end{subfigure}
	\caption[Universal Perspective examples]{Examples of distortion in Universal Perspective model with vignette effect.}
	\label{fig:universal perspective}
\end{figure}

\begingroup\allowdisplaybreaks
\begin{subequations}
\begin{align}
	&\begin{bmatrix}
		\vec u_s \\ \vec u_t
	\end{bmatrix}
	=
	\begin{cases}
		\left\Vert\begin{bmatrix}
			a & 1
		\end{bmatrix}\right\Vert\transpose, & \text{if\, $\Omega$ diagonal}
		\smallskip\\
		\begin{bmatrix}
			1 & \frac{1}{a}
		\end{bmatrix}\transpose, & \text{if\, $\Omega$ horizontal}
		\smallskip\\
		\frac{3}{4}\begin{bmatrix}
			a & 1
		\end{bmatrix}\transpose, & \text{if\, $\Omega$ horizontal 4$\times$3}
		\smallskip\\
		\begin{bmatrix}
			a & 1
		\end{bmatrix}\transpose, & \text{if\, $\Omega$ vertical}
	\end{cases}
	\tag{mapping vector}
\\[0.38em]
	&\begin{bmatrix}
		\vec f_x \smallskip\\ \vec f_y
	\end{bmatrix}
	=
	\begin{bmatrix}
		\vec f_s \\ \vec f_t \\ 1
	\end{bmatrix}
	\begin{bmatrix}
		2\vec u_s &0 &-\vec u_s \\
		0 &2\vec u_t &-\vec u_t
	\end{bmatrix}
\\[1.62em]
	&\left.
	\begin{aligned}
		R &= \Bigg|\begin{bmatrix}
			\vec f_x \smallskip\\
			\sqrt l\cdot\vec f_y
		\end{bmatrix}\Bigg|
		= \sqrt{\vec f_x^2+l\vec f_y^2}
	\\
		\theta &=
		\begin{cases}
			\arctan{ \big( R\cdot \tan\left( k\frac{\Omega}{2} \right) \big) } \div k,
			& \text{if } k>0 \\
			R \frac{\Omega}{2},
			& \text{if } k=0 \\
			\arcsin{ \big( R\cdot \sin\left( k\frac{\Omega}{2} \right) \big) } \div k,
			& \text{if } k<0
		\end{cases}
	\\[1em]
		\begin{bmatrix}
			\vec f'_x \smallskip\\
			\vec f'_y
		\end{bmatrix}
		&= \frac{\tan\theta}{R}
		\begin{bmatrix}
			\vec f_x \\
			\vec f_y\left(\frac{1-l}{s}+l\right)
		\end{bmatrix} \qed
	\end{aligned}
	\;\right\}
	\tag{perspective map}
\\[1.62em]
	&\begin{bmatrix}
		\vec f'_s \smallskip\\
		\vec f'_t
	\end{bmatrix}
	=
	\begin{bmatrix}
		\vec f'_x \smallskip\\
		\vec f'_y \smallskip\\
		1
	\end{bmatrix}
	\begin{bmatrix}
		\frac{1}{2\vec u_s\tan(\nicefrac{\Omega}{2})} &0 &\nicefrac{1}{2} \\
		0 &\frac{1}{2\vec u_t\tan(\nicefrac{\Omega}{2})} &\nicefrac{1}{2}
	\end{bmatrix}
	\qed
\end{align}
\end{subequations}
\endgroup
$k\in[-1,1]$ is the perspective-type scalar, $l\in[0,1]$ denotes cylindrical/spherical type factor, while $s\in[\nicefrac{4}{5},1]$ is the anamorphic correction scalar of non-spherical image. $\Omega\in (0, \pi)$ represents angle of view.
$\vec f_{st}\in[0,1]^2$ is the initial texture coordinate.
Vector $\vec f'_{xy}\in\real^2$ represents the view coordinate (preserves AOV),
while $\vec f'_{st}$ denotes final \emph{STMap} coordinate, which is normalized to square coordinates.
For more information on this formula, see listing \vref{lst:universal STMap}.
\begin{rem}
	Value of $l>1$ can be used to compensate for curved-display distortion.
\end{rem}

\begin{table}[h]
	\centerline{
	\fbox{\begin{tabular}[t]{r l}
		\multicolumn{2}{r}{\textbf{Perspective type}} \\
		\hline
		Rectilinear (gnomonic) & $k = 1$ \\
		\hline
		Stereographic & $k = \nicefrac{1}{2}$ \\
		\hline
		Equidistant & $k = 0$ \\
		\hline
		Equisolid & $k = -\nicefrac{1}{2}$ \\
		\hline
		Orthographic & $k = -1$
	\end{tabular}
	\begin{tabular}[t]{r l}
		\multicolumn{2}{r}{\textbf{Projection type}} \\
		\hline
		Spherical & $l = 1$ \\
		\hline
		Cylindrical & $l = 0$
	\end{tabular}
	\begin{tabular}[t]{c l}
		\multicolumn{2}{r}{\textbf{Default values}} \\
		\hline
		$\Omega^h$ & $\Rightarrow 90\degree$ \\
		\hline
		$k$ & $\Rightarrow 1$ \\
		\hline
		$l$ & $\Rightarrow 100\%$ \\
		\hline
		$s$ & $\Rightarrow 98\%$
	\end{tabular}}}

	\caption[Parameters of Universal Perspective]{Universal Perspective parameters and associated projection type.}
	\label{tab:universal perspective parameters}
\end{table}

\paragraph{Natural vignetting}
\label{sub:vignette}
occurs when visual sphere is projected onto an image plane with varying area of projection. Adding accurate vignetting to the digitally generated image increases visual experience. Additionally natural illumination falloff gives subconscious visual cue on geometry of projected image. Such cue is otherwise absent in motionless picture depicting space of unfamiliar objects to the viewer.
Below vignette formula is presented as a part of the universal perspective model described above.

\begin{subequations}
\begin{align}
	&\theta' = \theta\cdot\max\big\{|k|, \nicefrac{1}{2}\big\}
\\
	&v_s = \text{Lerp}\Big(
		\overbrace{\cos\theta'}^\text{cosine law},\quad
		\overbrace{\big(1+\tan^2\theta'\big)^{-1}}^\text{inverse-square law};\quad
		\big\{k+\nicefrac{1}{2}\big\}\cap[0,1]
	\Big)
	\tag{spherical}
\\
	&v_c = \underbrace{\left|\begin{bmatrix}
		\frac{\sin\theta}{R}\vec f_x &
		\frac{\sin\theta}{R}\vec f_y &
		\cos\theta
	\end{bmatrix}\right|^{-2}}_\text{inverse-square law}
	\tag{cylindrical}
\\[1em]
	&\left\{
	\begin{aligned}
		v &= v_s v_c \qed
	\\
		v' &= \text{gamma}\big(v,\,2.2\big)
	\end{aligned}
	\right.
\end{align}
\end{subequations}
Vignette mask $v$ is based upon \emph{inverse-square law} and a \emph{cosine law}  of illumination.
$v_s$ represents spherical perspective vignette and $v_c$, cylindrical projection vignette.
Gamma corrected mask $v'$ uses $\gamma=2.2$ (\emph{sRGB}).
Here function for gamma is defined as follows:
\begin{equation}
	\text{gamma}\big(w,\gamma\big) = w^{\nicefrac{1}{\gamma}}
\end{equation}
For more information on above equations see listing \vref{lst:universal STMap}.
\begin{table}[h]
	\centering
	\begin{tabular}{c l|l}
		\toprule
		\hfill\textbf{Projection type} & \hfill\textbf{Value range} & \hfill\textbf{Illumination law} \\ \hline
		orthographic\hfill$\leftrightarrow$\hfill equisolid &
		$k\in[-1,\nicefrac{1}{2}]$ &
		\emph{Cosine law} of illumination
		\\
		$\mapsfrom$ equidistant $\mapsto$ &
		$k\in(-\nicefrac{1}{2},\nicefrac{1}{2})$ &
		\emph{Cosine \& inverse-square law}
		\\
		stereographic $\leftrightarrow$ rectilinear &
		$k\in[\nicefrac{1}{2},1]$ &
		\emph{Inverse-square law} \\
		\bottomrule
	\end{tabular}

	\caption[Illumination laws and vignetting]{Laws of illumination falloff used in vignetting effect with Universal Perspective model. Sorted per projection type.}
	\label{tab:illumination law}
\end{table}
\begin{rem}
	Linear vignette mask $v$ can be added as the alpha channel of the rasterization map.
\end{rem}
\begin{rem}
	For $\gamma = 2$ and $k=1$, the result $v'$ conforms to $\cos^4$ or \textitquote{cosine fourth} law of illumination falloff; a vignette approximation for wide-angle lens picture.
\end{rem}

\begin{lstlisting}[float=p,
	label={lst:universal STMap},
	caption={
		[Universal Perspective STMap]
		Universal Perspective \emph{STMap} and natural vignetting mask in GLSL.
	}
]
// STMap generator in universal perspective
vec4 univSTMap(int FOV_type, int FOV, float k, float l, float s, vec2 texCoords, float aspect)
{
	vec2 mapping = (FOV_type==1)?
		vec2(1.0, 1.0/aspect) : // Horizontal FOV
		vec2(aspect, 1.0); // Vertical, diagonal, 4:3 FOV
	if (FOV_type==0) mapping = normalize(mapping); // Diagonal FOV
	if (FOV_type==2) mapping *= 0.75; // 4:3 FOV
	// Center coordinates and correct aspect
	texCoords = (2.0*texCoords-1.0)*mapping;

	// Get half AOV in radians
	float halfOmega = radians(FOV*0.5);
	// Get radius
	float R = length(vec2(texCoords.x, sqrt(l)*texCoords.y));

	float theta; // Incident angle
	if (k>0.0)      theta = atan(R*tan(k*halfOmega))/k;
	else if (k<0.0) theta = asin(R*sin(k*halfOmega))/k;
	else            theta = R*halfOmega;

	float vignette = 1.0;
	// Apply cylindrical vignette
	if (l!=1.0) // optimization
	{
		vec3 incident = vec3(texCoords*(sin(theta)/R), cos(theta));
		vignette /= dot(incident, incident); // Inverse square law
	}

	// Apply perspective transformation
	texCoords *= (tan(theta)/R);
	// Anamorphic correction of non-spherical image
	texCoords.y /= mix(s, 1.0, l);

	// Apply spherical vignette
	theta *= max(abs(k), 0.5); // k+- in [0.5,1]->[0,1]
	vignette *= mix(
		cos(theta), // Cosine law of illumination
		1.0/(pow(tan(theta), 2)+1.0), // Inverse square law
		clamp(k+0.5, 0.0, 1.0) // k in [-0.5,0.5]->[0,1]
	);

	// Map to square and corner
	texCoords = (texCoords/mapping)*(0.5/tan(halfOmega))+0.5;

	// Return STMap with linear vignette mask
	return vec4(texCoords, 0.0, vignette);
}
\end{lstlisting}

\subsubsection[Perspective Map conversion]{\emph{Perspective Map} to \emph{STMap}}
\label{sub:perspective map conversion}

\emph{Perspective Map} encodes pixel position in \emph{view space} as a unit vector of visual sphere.
It preserves angle of view, which contrary to \emph{STMap} can exceed 180\degree.
Therefore conversion between Perspective and \emph{STMap} is limited to views below 180\degree.

\begin{subequations}
\begin{align}
	\begin{bmatrix}
		\vec u_s \\ \vec u_t
	\end{bmatrix}
	&=
	\begin{cases}
		\left\Vert\begin{bmatrix}
			a & 1
		\end{bmatrix}\right\Vert\transpose, & \text{if\, $\Omega$ diagonal}
		\smallskip\\
		\begin{bmatrix}
			1 & \frac{1}{a}
		\end{bmatrix}\transpose, & \text{if\, $\Omega$ horizontal}
		\smallskip\\
		\frac{3}{4}\begin{bmatrix}
			a & 1
		\end{bmatrix}\transpose, & \text{if\, $\Omega$ horizontal 4$\times$3}
		\smallskip\\
		\begin{bmatrix}
			a & 1
		\end{bmatrix}\transpose, & \text{if\, $\Omega$ vertical}
	\end{cases}
	\tag{mapping vector}
\\[0.62em]
	\begin{bmatrix}
		\vec f_s \\
		\vec f_t
	\end{bmatrix}
	&=
	\overbrace
	{\begin{bmatrix}
		\hat G_x \div \max\{z_n,\hat G_z\} \\
		\hat G_y \div \max\{z_n,\hat G_z\} \\
		1
	\end{bmatrix}}^\text{planar projection}
	\begin{bmatrix}
		\frac{\cot\nicefrac{\Omega}{2}}{2\vec u_s} &0 & \nicefrac{1}{2} \\
		0 &\frac{\cot\nicefrac{\Omega}{2}}{2\vec u_t} & \nicefrac{1}{2}
	\end{bmatrix}
	\qed
\\
	m &= \text{pixStep}\big(\hat G_z-z_n\big)
\end{align}
\end{subequations}
Here $\vec f_{st}$ represents \emph{STMap} coordinate. $a=\nicefrac{\textit{width}}{\textit{height}}$ denotes aspect ratio, while $\hat G\in[0,1]$ is the perspective-map unit vector.
Scalar $z_n\in(0,1)$ describes near-clip plane, with $m$ as optional bounds mask.
This conversion method is limited to angles $\Omega\in(0, \pi)$, which for FOV is between 1\degree$\leftrightarrow$179\degree. See listing \vref{lst:perspective map convert} for more information.

\begin{lstlisting}[float=p,
	label={lst:perspective map convert},
	caption={
		[Perspective Map to STMap]
		\emph{STMap} $\vec f$ coordinates from \emph{Perspective Map} vector $\hat G$, with FOV scaling in GLSL.
	}
]
vec4 Persp2STMap(vec3 sphVec, int FOV, int FOV_type, float aspect)
{
	// Plane projection with normalization
	vec2 STMap = (0.5/tan(radians(FOV*0.5))/sphVec.z)*sphVec.xy;

	// Diagonal FOV
	if (FOV_type==0) STMap *= inversesqrt(aspect*aspect+1.0);
	// Horizontal FOV
	if (FOV_type==1) STMap.y *= aspect;
	// Vertical (3) or diagonal or 4:3 FOV
	else STMap.x /= aspect;
	// 4:3 FOV
	if (FOV_type==2) STMap *= 4.0/3.0;

	// Get bounds mask
	vec2 bounds = abs(STMap)-0.5;
	mat2 DelMtx = transpose(mat2(dFdx(bounds), dFdy(bounds)));
	bounds = 1.0-clamp(inversesqrt(vec2(
		dot(DelMtx[0], DelMtx[0]),
		dot(DelMtx[1], DelMtx[1])
	))*bounds, 0.0, 1.0);

	return vec4(STMap+0.5, 0.0, min(bounds.x, bounds.y));
}
\end{lstlisting}

\subsection[Lens distortion STMap]{Lens distortion \emph{STMap}}
\label{sub:lens distortion}

Below presented is lens-distortion division model based on \emph{Brown-Conrady}'s,\supercite{Wang2008LensDistortion,Fitzgibbon2001LensDivision} but with angle-of-view preservation (normalization).
While AOV normalization makes correction by hand more difficult, it preserves picture's parameters.

\begin{subequations}
\begin{align}
	\begin{bmatrix}
		\vec u_s \\ \vec u_t
	\end{bmatrix}
	&=
	\begin{cases}
		\left\Vert
		\begin{bmatrix}
			a &1
		\end{bmatrix}
		\right\Vert,
		&\text{if AOV diagonal \ (full-frame)}
	\smallskip\\
		\begin{bmatrix}
			1 &\nicefrac{1}{a}
		\end{bmatrix},
		&\text{if AOV horizontal \ (cropped circle)}
	\smallskip\\
		\frac{3}{4}
		\begin{bmatrix}
			a &1
		\end{bmatrix},
		&\text{if AOV 4$\times$3 \ (cropped)}
	\smallskip\\
		\begin{bmatrix}
			a &1
		\end{bmatrix},
		&\text{if AOV vertical}
	\end{cases}
	\tag{mapping vector}
\\
	\begin{bmatrix}
		\vec f_x \\ \vec f_y
	\end{bmatrix}
	&=
	\begin{bmatrix}
		\vec f_s \\ \vec f_t \\ 1
	\end{bmatrix}
	\begin{bmatrix}
		2\vec u_s &0 &-\vec u_s \\
		0 &2\vec u_t &-\vec u_t \\
		0 &0 &1
	\end{bmatrix}
	\underbrace
	{\begin{bmatrix}
		1 &0 &-c_1 \\
		0 &1 &-c_2
	\end{bmatrix}}_\text{cardinal offset \textnumero1}
	\tag{View space}
\\
	r^2 &= \vec f^2_x + \vec f^2_y
\\
	\begin{bmatrix}
		\vec f'_x \\ \vec f'_y
	\end{bmatrix}
	&=
	\begin{bmatrix}
		\vec f_x \\ \vec f_y
	\end{bmatrix}
	+
	r^2
	\begin{bmatrix}
		q_1 \\ q_2
	\end{bmatrix} \qed
	\tag{decentering}
\\
	\begin{bmatrix}
		\vec f''_x \\ \vec f''_y
	\end{bmatrix}
	&=
	\begin{bmatrix}
		\vec f'_x \\ \vec f'_y
	\end{bmatrix}
	+
	\begin{bmatrix}
		\vec f'_x \\ \vec f'_y
	\end{bmatrix}
	\Bigg(
	\begin{bmatrix}
		p_1 \\ p_2
	\end{bmatrix}
	\cdot
	\begin{bmatrix}
		\vec f'_x \\ \vec f'_y
	\end{bmatrix}
	\Bigg) \qed
	\tag{thin prism}
\\
	\begin{bmatrix}
		\vec f'''_x \\ \vec f'''_y
	\end{bmatrix}
	&=
	\begin{bmatrix}
		\vec f''_x \\ \vec f''_y
	\end{bmatrix}
	\frac
	{\overbrace{1+k_1+k_2+\dots+k_n}^\text{bounds normalization}}
	{1+k_1r^2+k_2r^4+\dots+k_nr^{(2n)}} \qed
	\tag{radial}
\\[0.62em]
	\begin{bmatrix}
		\vec f'_s \\ \vec f'_t
	\end{bmatrix}
	&=
	\begin{bmatrix}
		\vec f'''_x \\ \vec f'''_y \\ 1
	\end{bmatrix}
	\underbrace
	{\begin{bmatrix}
		1 &0 &c_1 \\
		0 &1 &c_2 \\
		0 &0 &1
	\end{bmatrix}}_\text{cardinal offset \textnumero2}
	\begin{bmatrix}
		\frac{1}{2\vec u_s} &0 &\nicefrac{1}{2} \\
		0 &\frac{1}{2\vec u_t} &\nicefrac{1}{2}
	\end{bmatrix} \qed
	\tag{\emph{STMap} space}
\end{align}
\end{subequations}
Here $a=\nicefrac{\textit{width}}{\textit{height}}$ denotes aspect ratio. $\vec f_{st}\in[0,1]^2$ is the initial texture coordinate, with $\vec f'_{st}\in\real^2$ as the lens-transformed texture coordinate (see listing \vref{lst:lens distortion} for more information).
\begin{table}[h]
	\centering
	\begin{tabular}{r l l}
		\toprule
		$\{k_1,k_2,\dotsm,k_n\}$ & $\in[-\nicefrac{2}{5},\nicefrac{2}{5}]$ & Radial distortion coefficients \\
		$\{p_1,p_2\}$ & $\in[-\nicefrac{1}{5},\nicefrac{1}{5}]$            & Thin prism distortion coefficients \\
		$\{q_1,q_2\}$ & $\in[-\nicefrac{1}{10},\nicefrac{1}{10}]$          &
		Decentering coefficients \\
		$\{c_1,c_2\}$ & $\in[-\nicefrac{1}{5},\nicefrac{1}{5}]$            & Cardinal offset coefficients \\
		\bottomrule
	\end{tabular}

	\caption[Lens distortion parameters]{Lens distortion parameters with suggested value ranges.}
	\label{tab:lens parameters}
\end{table}

\begin{figure}[H]
	\centering
	\begin{subfigure}{60pt}
		\centerline{\includegraphics[width=60pt]{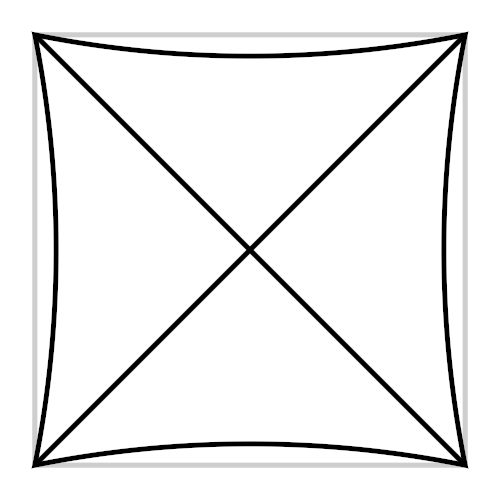}}
		\caption{$k_1=0.2$}
	\end{subfigure}\hspace{26pt}
	\begin{subfigure}{60pt}
		\centerline{\includegraphics[width=60pt]{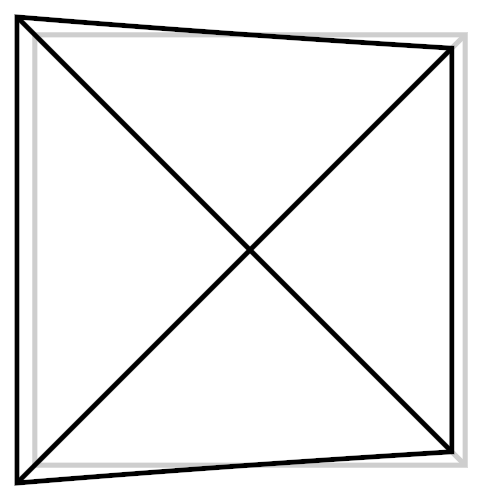}}
		\caption{$p_1=0.1$}
	\end{subfigure}\hspace{25pt}
	\begin{subfigure}{60pt}
		\centerline{\includegraphics[width=60pt]{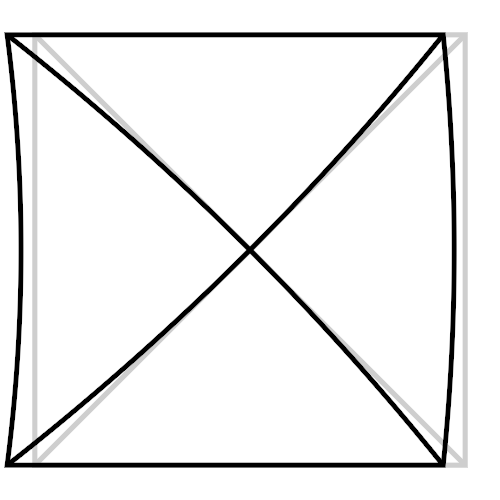}}
		\caption{$q_1=0.08$}
	\end{subfigure}\hspace{22.5pt}
	\begin{subfigure}{60pt}
		\centerline{\includegraphics[width=60pt]{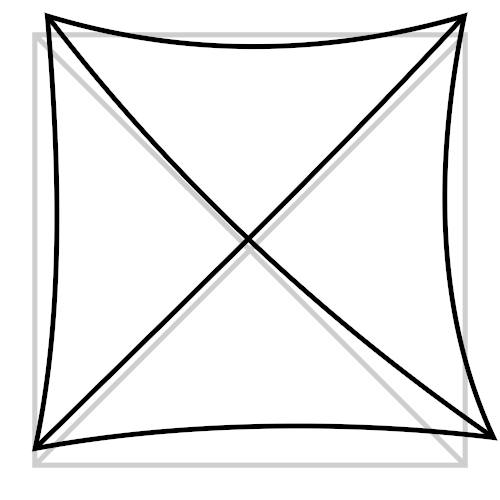}}
		\caption{}
	\end{subfigure}
\\[1em]
	\begin{subfigure}{60pt}
		\centerline{\includegraphics[width=60pt]{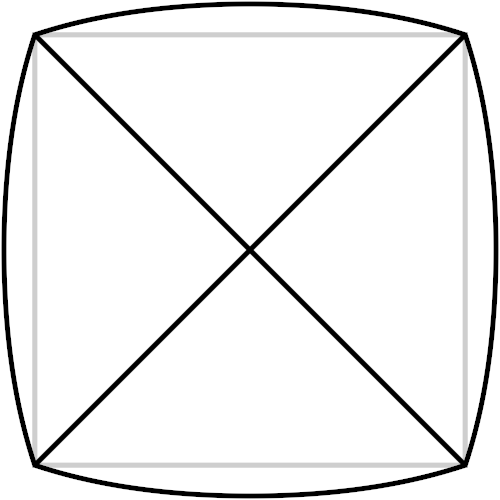}}
		\caption{$k_2=-0.2$}
	\end{subfigure}\hspace{26pt}
	\begin{subfigure}{60pt}
		\centerline{\includegraphics[width=60pt]{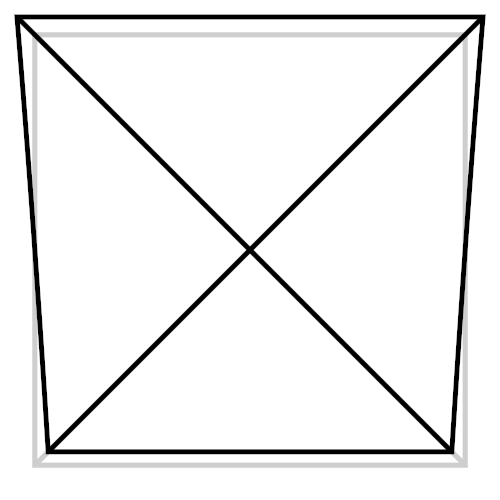}}
		\caption{$p_2=-0.1$}
	\end{subfigure}\hspace{25pt}
	\begin{subfigure}{60pt}
		\centerline{\includegraphics[width=60pt]{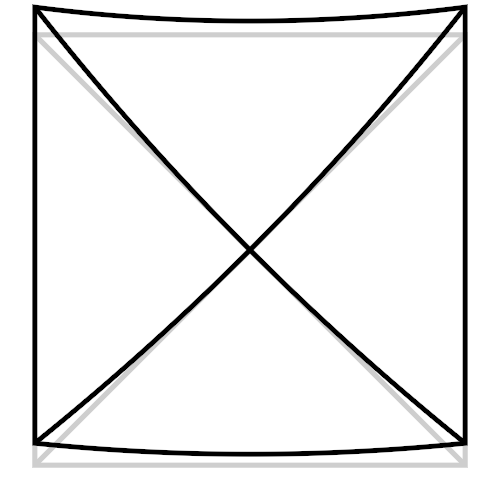}}
		\caption{$q_2=-0.08$}
	\end{subfigure}\hspace{22.5pt}
	\begin{subfigure}{60pt}
		\centerline{\includegraphics[width=60pt]{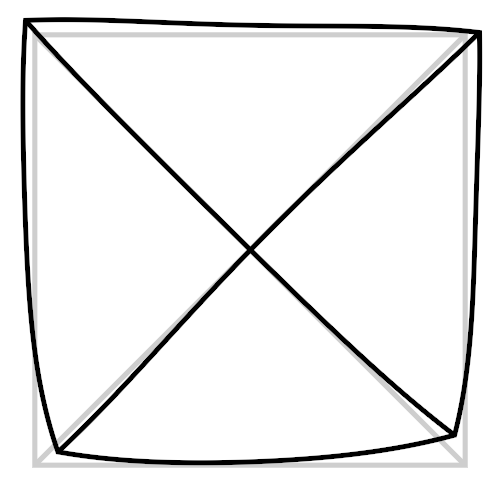}}
		\caption{}
	\end{subfigure}
	\caption[Lens distortion examples]
	{Lens distortion examples, with full-frame (diagonal) AOV normalization. Sub-figure (d) represents $k=[0.2\quad0]$, $p=[0.04\quad-0.04]$, $q=[-0.04\quad-0.1]$ and $c=[-0.04\quad0.2]$. While sub-figure (h) represent $k=[0.2\quad-0.2]$, $p=[0.04\quad-0.04]$, $q=[-0.04\quad0.0]$ and $c=[-0.04\quad0.2]$.}
	\label{fig:lens distortion}
\end{figure}

\begin{rem}
	Note that extreme lens distortions, like fish-eye view should be additionally mapped to another projection, before the lens-correction. For example Universal Perspective mapping can be used, which can be found in subsection \vref{sub:universal perspective}.
\end{rem}

\begin{lstlisting}[float=p,
	label={lst:lens distortion},
	caption={
		[Lens distortion STMap]
		Lens distortion \emph{STMap} coordinates with FOV preservation, in GLSL.
	}
]
// Radial distortion-division model with normalization
vec2 lensRadial(vec2 texCoords, vec2 k)
{
	float R2 = dot(texCoords, texCoords); // Radius squared
	return (1+k[0]+k[1])/((1+k[0]*R2)+k[1]*pow(R2, 2))*texCoords;
}
// Thin prism distortion model
vec2 lensThinPrism(vec2 texCoords, vec2 p)
{ return texCoords*dot(p, texCoords)+texCoords; }
// Decentering distortion model
vec2 lensDecentering(vec2 texCoords, vec2 q)
{
	float R2 = dot(texCoords, texCoords); // Radius squared
	return R2*q+texCoords;
}
// Lens distortion texture coordinates
vec2 lensDistort(vec2 k, vec2 p, vec2 q, vec2 c, vec2 texCoords, int type, float aspect)
{
	vec2 mapping = type==1? // Horizontal or Vertical AOV?
		vec2(1.0, 1.0/aspect) : vec2(aspect, 1.0);
	if (type==0) mapping = normalize(mapping); // Diagonal AOV
	if (type==2) mapping *= 0.75; // 4:3 AOV
	texCoords = (2.0*texCoords-1.0)*mapping; // Apply normalization

	texCoords -= c; // Cardinal offset No.1
	texCoords = lensDecentering(texCoords, q); // Decentering correction
	texCoords = lensThinPrism(texCoords, p); // Thin prism correction
	texCoords = lensRadial(texCoords, k); // Radial correction
	texCoords += c; // Cardinal offset No.2

	return 0.5*(texCoords/mapping)+0.5;
}
// Lens distortion STMap from texture coordinates
vec4 lensDistortSTMap(vec2 texCoords)
{
	vec2 bounds = abs(texCoords-0.5)-0.5;
	mat2 DelMtx = transpose(mat2(dFdx(bounds), dFdy(bounds)));
	bounds = 1.0-clamp(inversesqrt(vec2(
		dot(DelMtx[0], DelMtx[0]),
		dot(DelMtx[1], DelMtx[1])
	))*bounds, 0.0, 1.0);

	return vec4(texCoords, 0.0, min(bounds.x, bounds.y));
}
\end{lstlisting}

\subsection[Rasterization with Perspective Map]{Rasterization with lookup spherical coordinates}
\label{sub:lookup spherical rasterization}

Rasterization process with \emph{Perspective Map} does not differ much from \emph{STMap} rasterization.
Rasterization matrix uses normalized vector in denominator, as \emph{Perspective Map} encodes a unit-sphere. Depth pass is replaced by a distance pass. Perspective-correct barycentric coordinates are obtained using the same method as previously (see equation \vref{eq:barycentric correct}), but with distance.

\begin{equation}
	\begin{bmatrix}
		\chi_{1,1} & \chi_{1,2} & \chi_{1,3} \\
		\chi_{2,1} & \chi_{2,2} & \chi_{2,3} \\
		\chi_{3,1} & \chi_{3,2} & \chi_{3,3}
	\end{bmatrix}
	=
	\begin{bmatrix}
		\bigg[\frac{\vec B\cross\vec C}{\Vert\hat A\Vert\cdot(\vec B\cross\vec C)}\bigg]\transpose 
	\\[0.62em]
		\bigg[\frac{\vec C\cross\vec A}{\Vert\hat B\Vert\cdot(\vec C\cross\vec A)}\bigg]\transpose 
	\\[0.62em]
		\bigg[\frac{\vec A\cross\vec B}{\Vert\hat C\Vert\cdot(\vec A\cross\vec B)}\bigg]\transpose 
	\end{bmatrix}
	\qquad
	\begin{matrix}
		\text{edge }f(a) \\[1.38em]
		\text{edge }f(b) \\[1.38em]
		\text{edge }f(c)
	\end{matrix}
\end{equation}
$\chi\in\real^{3\times3}$ is the \emph{Perspective Map} rasterization matrix. Edge functions (represented by matrix $\chi$ rows) are converted into barycentric functions by the dot-product in denominator. See listing \vref{lst:perspective map rasterization} for more information.

\begin{subequations}
\begin{align}
	\begin{bmatrix}
		\vec\lambda_s \\
		\vec\lambda_t \\
		\vec\lambda_p
	\end{bmatrix}
	&=
	\begin{bmatrix}
		\hat G_x \\
		\hat G_y \\
		\hat G_z
	\end{bmatrix}
	\begin{bmatrix}
		\chi_{1,1} & \chi_{1,2} & \chi_{1,3} \\
		\chi_{2,1} & \chi_{2,2} & \chi_{2,3} \\
		\chi_{3,1} & \chi_{3,2} & \chi_{3,3}
	\end{bmatrix} \qed
	\tag{Spherical barycentric}
\\
	\begin{bmatrix}
		\vec\Xi_1 \smallskip\\
		\vec\Xi_2 \smallskip\\
		\vec\Xi_3
	\end{bmatrix}
	&=
	\left\{
	\vphantom{\begin{bmatrix}
		\vec\lambda_s \smallskip\\
		\vec\lambda_t \smallskip\\
		\vec\lambda_p
	\end{bmatrix}}
	\right.
	\overbrace{\frac{1}{2}+
	\begin{bmatrix}
		{|\vec{\nabla\lambda_s}|}^{-1}
	\smallskip\\
		{|\vec{\nabla\lambda_t}|}^{-1}
	\smallskip\\
		{|\vec{\nabla\lambda_p}|}^{-1}
	\end{bmatrix}}^\text{anti-aliasing}
	\left.
	\begin{bmatrix}
		\vec\lambda_s \smallskip\\
		\vec\lambda_t \smallskip\\
		\vec\lambda_p
	\end{bmatrix}
	\right\}
	\cap [0,1]
\\
	\Lambda &= \prod^3_{i=1} \vec \Xi_i \qed
\end{align}
\end{subequations}
Here $\vec\lambda\in\real^3$ represents spherical-barycentric vector, with $\hat G\in[0,1]^3$ as the spherical-incident vector. $\vec\Xi\in[0,1]^3$ is an RMAA half-space triangle with $\Lambda\in[0,1]$ as the triangle coverage mask.
See more in listing \vref{lst:perspective map rasterization}.

\begin{equation}
	\Lambda' =
		\big(\vec\lambda_s\geq0\big)\land
		\big(\vec\lambda_t\geq0\big)\land
		\big(\vec\lambda_p\geq0\big) \mapsto \{0,1\}
	\qed
\end{equation}
Aliased (binary) coverage mask $\Lambda'\in\{0,1\}$ is obtained by conjunction.

Spherical-barycentric vector $\vec\lambda$ requires perspective-correction to be used as vertex-data interpolation weight.
To correct barycentric vector, first distance pass $\vec f_d$ is obtained using inverse of the dot-product between spherical-barycentric coordinate $\vec\lambda$ and the inversed vertex-distance vector.
Result distance pass is perspective-correct. From it, corrected barycentric vector $\vec\lambda'$ is produced.
\begin{subequations}
\begin{align}
	\vec f_d
	&=
	\left(
	\begin{bmatrix}
		\vec\lambda_s \\
		\vec\lambda_t \\
		\vec\lambda_p
	\end{bmatrix}
	\cdot
	\begin{bmatrix}
		|\vec A|^{-1} \\
		|\vec B|^{-1} \\
		|\vec C|^{-1}
	\end{bmatrix}
	\right)^{-1}
	\\
	\begin{bmatrix}
		\vec\lambda'_s \\
		\vec\lambda'_t \\
		\vec\lambda'_p
	\end{bmatrix}
	&=
	\vec f_d
	\begin{bmatrix}
		\vec\lambda_s \\
		\vec\lambda_t \\
		\vec\lambda_p
	\end{bmatrix}
	\begin{bmatrix}
		|\vec A|^{-1} \\
		|\vec B|^{-1} \\
		|\vec C|^{-1}
	\end{bmatrix}
\end{align}
\end{subequations}
Scalar $\vec f_d\in\real_{>0}$ represent interpolated distance pass, which replaces depth pass. Vector $\vec\lambda'\in\real^3$ is the perspective-correct barycentric vector.
See listing \vref{lst:perspective map rasterization} for more information.

\begin{lstlisting}[float=p,
	label={lst:perspective map rasterization},
	caption={
		[Perspective Map rasterization]
		Rasterization with \emph{Perspective Map} in GLSL.
		For the $\textbit{pixStep}\big(\vec\lambda\big)$ function definition see listing \vref{lst:anti-aliasing}.
	}
]
#define RMAA true

// Get barycentric-edge function
vec3 getEdgeFunc(vec3 P0, vec3 P1, vec3 P2)
{
	vec3 edgeFunc = cross(P0, P1);
	// Convert to barycentric function
	return edgeFunc/dot(normalize(P2), edgeFunc);
}
// Get barycentric rasterization matrix
mat3 getEdgeMtx(mat3 polyg)
{
	return mat3(
		getEdgeFunc(polyg[1], polyg[2], polyg[0]), // Line a
		getEdgeFunc(polyg[2], polyg[0], polyg[1]), // Line b
		getEdgeFunc(polyg[0], polyg[1], polyg[2])  // Line c
	);
}
// Get vector of inverse distance of each triangle point
vec3 getRcpDistance(mat3 polyg)
{
	return vec3(
		inversesqrt(dot(polyg[0], polyg[0])),
		inversesqrt(dot(polyg[1], polyg[1])),
		inversesqrt(dot(polyg[2], polyg[2]))
	);
}
// Returns distance and perspective-correct barycentric coordinates
float correctBarycentric(inout vec3 barycentric, vec3 rcpVrtxDist)
{
	float distance = 1.0/dot(rcpVrtxDist, barycentric);
	barycentric *= distance*rcpVrtxDist; return distance; // Output
}
// Rasterize triangle coverage and linear-barycentric coordinates
vec4 rasterize(vec3 incident, mat3 rasterMtx)
{
	vec3 barycentric = incident*rasterMtx;
	vec3 halfSpace = RMAA? pixStep(barycentric) : step(0.0, barycentric);
	// Return barycentric coordinates and coverage mask
	return vec4(barycentric, halfSpace.r*halfSpace.g*halfSpace.b);
}
\end{lstlisting}

\section{Requirements and recommendations}
\label{sec:recommendations}

\begin{table}[H]
	\centering
	\begin{tabular}{l l l}
		\toprule
		\textbf{RMAA only} & \textbf{\emph{STMap} only} &\textbf{\emph{STMap} \& RMAA} \\ \midrule
		\textbullet\ Rasterizer & \textbullet\ Rasterizer & \textbullet\ Rasterizer \\
		\textbullet\ Output merger & \textbullet\ Bounding box & \textbullet\ Bounding box \\
		\textbullet\ Front-to-back order & & \textbullet\ Output merger \\
			& & \textbullet\ Front-to-back order \\[3pt]
		\toprule
		\textbf{Perspective-map only} & \multicolumn{2}{l}{\textbf{Perspective-map \& RMAA}} \\ \midrule
		\textbullet\ Rasterizer & \textbullet\ Rasterizer \\
		\textbullet\ Bounding box & \textbullet\ Bounding box \\
		\textbullet\ Culling space & \textbullet\ Culling space \\
			& \textbullet\ Output merger \\
			& \textbullet\ Front-to-back order \\
		\bottomrule
	\end{tabular}
	\caption[Required changes in rasterization pipeline]{Changes required in standard rasterization-model pipeline, arranged by features.}
	\label{tab:requirements}
\end{table}

\paragraph{Rasterization \emph{STMap}}
with barrel distortion should utilize diagonal AOV normalization (denoted $\Omega^d$), which is the widest AOV.
Diagonal normalization avoids visibility of out-of-bounds coordinates.
In contrary, with pincushion distortion, \emph{STMap} should utilize the shortest AOV (for panoramic picture it is a vertical AOV, denoted $\Omega^v$).

\emph{STMap} preserves floating-point precision but not camera properties. Therefore it is recommended to include camera information within \emph{STMap} file.

\begin{rem}
	Initial resolution of the \emph{STMap} and \emph{Perspective Map} must be an even number, both vertically and horizontally. This is due to way, in which rasterization maps are calculated. Result vector texture can be later scaled to desired format.
\end{rem}

\subsection[File naming convention]{Rasterization map naming convention}
\label{sub:file naming convention}

\begin{table}[h]
	\begin{subtable}{0.5\textwidth}
		\centering
		\fbox{\begin{tabular}{r|r}
			\textbf{Symbol} & \textbf{FOV type} \\ \hline
			\texttt{h} & horizontal FOV \\
			\texttt{v} & vertical FOV \\
			\texttt{d} & diagonal FOV \\
			\texttt{4x3h}  & 4$\times$3--clipped FOV \\
			\texttt{16x9h} & 16$\times$9--clipped FOV \\
		\end{tabular}}
		\caption{Camera field of view.}
	\end{subtable}
	\begin{subtable}{0.5\textwidth}
		\centering
		\fbox{\begin{tabular}{r|r}
			\textbf{Symbol} & \textbf{Parameter type} \\ \hline
			\texttt{k} & projection type \\
			\texttt{l} & cylindrical factor \\
			\texttt{s} & anamorphic correction \\
		\end{tabular}}
		\caption{Universal Perspective parameters.}
	\end{subtable}
	\\[0.382em]
	\begin{subtable}{\textwidth}
		\centering
		\fbox{\begin{tabular}{r|r|c|l}
			\textbf{Symbol} & \textbf{Layer type} & \multicolumn{2}{r}{\textbf{Format}} \\ \hline
			\texttt{Pm}  & Perspective Map & \textbit{vec3} & \textbit{RGB32} \\
			\texttt{St}  & STMap & \textbit{vec2} & \textbit{RG32} \\
			\texttt{uSt} & undistort STMap & \textbit{vec2} & \textbit{RG32} \\
			\texttt{P}   & parallax map & \textbit{float} & \textbit{R32} \\
			\texttt{V}   & vignetting mask & \textbit{float} & \textbit{R8} \\
			\texttt{M}   & bounds mask & \textbit{float} & \textbit{R8} \\
			\texttt{PmV} & Perspective and vignetting mask & \textbit{vec4} & \textbit{RGBA32} \\
			\texttt{PmP} & Perspective and parallax map & \textbit{vec4} & \textbit{RGBA32} \\
			\texttt{PmM} & Perspective and bounds mask & \textbit{vec4} & \textbit{RGBA32} \\
			\texttt{StV} & STMap and vignetting mask & \textbit{vec3} & \textbit{RGB32} \\
			\texttt{StP} & STMap and parallax map & \textbit{vec3} & \textbit{RGB32} \\
			\texttt{StM} & STMap and bounds mask & \textbit{vec3} & \textbit{RGB32} \\
		\end{tabular}}
		\caption{Layers data.}
	\end{subtable}

	\caption[File naming convention]{File naming-convention glossary, of rasterization maps.}
	\label{tab:naming convention}
\end{table}

\begin{figure}[H]
	\large
	\begin{equation*}
		\underbrace{\texttt{AnamorphicWide}}_\text{description}
		\texttt{\_}
		\underbrace{\texttt{(Pm\_St\_P\_V)}}_\text{included layers}
		\texttt{\_}
		\underbrace{\texttt{(d140\_k0\_l0.62\_s0.98)}}_\text{perspective properties}
		\texttt{.exr}
	\end{equation*}
	\begin{equation*}
		\underbrace{\texttt{iDome}}_\text{description}
		\texttt{\_}
		\underbrace{\texttt{PmV}}_\text{data}
		\texttt{.exr}
	\end{equation*}

	\caption[Naming convention example]{Examples of naming convention for rasterization-map files.}
	\label{fig:naming convention}
\end{figure}

\section{Future work and extensions}
\label{sec:future}

Future extensions of \emph{Perspective Map} rasterization can incorporate world-position pass for screen-space shadowing with penumbra.
The result could rasterize final-pixel soft/hard-shadows and mimic penumbra effect, without intermediate re-projection.
Additionally render-region (or bounding box) evaluation should be further investigated as here, solution was only suggested.
Hardware-optimized version of the rasterizer should be finalized with evaluation of performance statistics.

\section[Conclusion]{Conclusion}
\label{sec:conclusion}

RMAA is a really fast and cheap anti-aliasing technique which outperforms other available solutions in visual fidelity. Unfortunately it requires fundamental change in Graphics Processing Unit (GPU) design, which is responsible of rasterization.

RMAA requires front-to-back order of rasterization. This sorting of data reduces overdraw but may be difficult to implement.
With increasing monitor resolutions, anti-aliasing may in some case become unnecessary.

Ordered rasterization is required only with RMAA technique, meaning that
rasterization with \emph{STMap} or \emph{Perspective Map} without RMAA can simply utilize Z-buffer test (or D-buffer with the latter), for solving the hidden-surface problem.

\begin{table}[h]
	\begin{subtable}{\textwidth}
		\;
		\fbox{\begin{tabular}{l|l}
			\textbf{Achilles heel} & \textbf{Unique points} \\
			\hline
			\textbullet\ Primitive sorting & \textbullet\ Very fast AA \\
			\textbullet\ GPU architecture interference & \textbullet\ High-fidelity of edges \\
		\end{tabular}}
		\caption{RMAA rasterization} 
	\end{subtable}
	\\[0.382em]
	\begin{subtable}{\textwidth}
		\;
		\fbox{\begin{tabular}{l|l}
			\textbf{Achilles heel} & \textbf{Unique points} \\
			\hline
			\textbullet\ Render region determination &\textbullet\ Unlimited distortion \\
			\textbullet\ \emph{STMap} texture sampling & \textbullet\ High-precision values \\
			\textbullet\ GPU architecture interference & \textbullet\ Industry-backed
		\end{tabular}}
		\caption{\emph{STMap} rasterization without AA} 
	\end{subtable}
	\\[0.382em]
	\begin{subtable}{\textwidth}
		\;
		\fbox{\begin{tabular}{l|l}
			\textbf{Achilles heel} & \textbf{Unique points} \\
			\hline
			\textbullet\ Render region determination &\textbullet\ Unlimited distortion \\
			\textbullet\ Perspective texture sampling & \textbullet\ Unlimited field of view \\
			\textbullet\ GPU architecture interference & \textbullet\ Unlimited views from a single-point
		\end{tabular}}
		\caption{\emph{Perspective Map} rasterization without AA} 
	\end{subtable}
	\caption[Summary of weak and strong points]{Summary of weak and strong points of RMAA technique, \emph{STMap} and \emph{Perspective Map} rasterization.}
	\label{tab:summary}
\end{table}







\pagebreak
\addcontentsline{toc}{section}{References}
\label{sec:references}
\printbibliography


\begin{figure}[b]
	\phantomsection
	\addcontentsline{toc}{section}{License notice}
	\hypertarget{license}{
		\footnotesize\noindent\copyright\ {\the\year} Jakub Maksymilian Fober (the Author).\quad The Author provides this document (the Work) under the Creative Commons CC BY-NC-ND 3.0 license available online. To view a copy of this license, visit \href{\license}{\license} or send a letter to Creative Commons, PO Box 1866, Mountain View, CA 94042, USA. The Author further grants permission for reuse of \hyperref[pg:title]{images and text from the first page} of the Work, provided that the reuse is for the purpose of promoting and/or summarizing the Work in scholarly venues and that any reuse is accompanied by a scientific citation to the Work.\vspace{4pt} \\
		\centerline{
			\href{\license}{\includegraphics{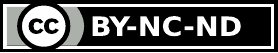}}
		}
	}
\end{figure}

\end{document}